\documentclass[longauth]{aa} 
\usepackage{graphicx}

\usepackage{natbib}
\bibpunct{(}{)}{;}{a}{}{,} 

\usepackage{txfonts}

\begin{document}

\title{
The Gaia-ESO Survey\thanks{Based on the data obtained at ESO telescopes under programme 188.B-3002
 (the public Gaia-ESO spectroscopic survey, PIs Gilmore \& Randich) and on the archive data of the programme 083.D-0671}: Stellar content and elemental abundances in the massive cluster NGC\,6705
}

   \subtitle{}

\author{
T. Cantat-Gaudin\inst{\ref{UNIPD},\ref{OAPD}}
\and
A. Vallenari\inst{\ref{OAPD}}
\and
S. Zaggia \inst{\ref{OAPD}}
\and
A. Bragaglia\inst{\ref{OABO}}
\and
R. Sordo\inst{\ref{OAPD}}
\and
J.~E. Drew\inst{\ref{HERTS}}
\and
J. Eisloeffel\inst{\ref{Tautenburg}}
\and
H.~J. Farnhill\inst{\ref{HERTS}}
\and
E. Gonzalez-Solares\inst{\ref{CASU}}
\and
R. Greimel\inst{\ref{IGAM}}
\and
M.~J. Irwin\inst{\ref{CASU}}
\and
A. Kupcu-Yoldas\inst{\ref{CASU}}
\and
C. Jordi\inst{\ref{AMB}}
\and
R. Blomme\inst{\ref{BRUSSEL}}
\and
L. Sampedro\inst{\ref{IAA}}
\and
M.~T. Costado\inst{\ref{IAA}}
\and
E. Alfaro\inst{\ref{IAA}}
\and
R. Smiljanic\inst{\ref{GARCHING},\ref{TORUN}}
\and
L. Magrini\inst{\ref{ARCETRI}}
\and
P. Donati \inst{\ref{OABO},\ref{UNIBO}}
\and
E.~D. Friel\inst{\ref{INDIANA}}
\and
H. Jacobson\inst{\ref{MIT}}
\and
U. Abbas\inst{\ref{OATO}}
\and
D. Hatzidimitriou\inst{\ref{ATHENS}}
\and
A. Spagna\inst{\ref{OATO}}
\and
A. Vecchiato\inst{\ref{OATO}}
\and
L. Balaguer-Nunez\inst{\ref{AMB}}
\and
C. Lardo\inst{\ref{OABO}}
\and
M. Tosi\inst{\ref{OABO}}
\and
E. Pancino\inst{\ref{OABO}}
\and
A. Klutsch\inst{\ref{CATANIA}}
\and
G. Tautvaisiene\inst{\ref{VILNIUS}}
\and
A. Drazdauskas\inst{\ref{VILNIUS}}
\and
E. Puzeras\inst{\ref{VILNIUS}}
\and
F. {Jim{\'e}nez-Esteban}\inst{\ref{SVO}}
\and
E. Maiorca\inst{\ref{ARCETRI}}
\and
D. Geisler\inst{\ref{CONCEPCION}}
\and
I. San Roman\inst{\ref{CONCEPCION}}
\and
S. Villanova\inst{\ref{CONCEPCION}}
\and
G. Gilmore\inst{\ref{IOA}}
\and
S. Randich\inst{\ref{ARCETRI}}
\and
T. Bensby\inst{\ref{LUND}}
\and
E. Flaccomio\inst{\ref{PALERMO}}
\and
A. Lanzafame\inst{\ref{CATANIA}}
\and
A. Recio-Blanco\inst{\ref{NICE}}
\and
F. Damiani\inst{\ref{PALERMO}}
\and
A. Hourihane\inst{\ref{IOA}}
\and
P. Jofr\'e\inst{\ref{IOA}}
\and
P. de Laverny\inst{\ref{NICE}}
\and
T. Masseron\inst{\ref{IOA}}
\and
L. Morbidelli\inst{\ref{ARCETRI}}
\and
L. Prisinzano\inst{\ref{PALERMO}}
\and
G.~G. Sacco\inst{\ref{ARCETRI}}
\and
L. Sbordone\inst{\ref{MAS},\ref{PUC}}
\and 
C.~C. Worley\inst{\ref{IOA}}
}

\institute{
Dipartimento di Fisica e Astronomia, Universit\`a di Padova, vicolo Osservatorio 3, 35122 Padova, Italy\label{UNIPD}
\and
INAF-Osservatorio Astronomico di Padova, vicolo Osservatorio 5, 35122 Padova, Italy\label{OAPD}
\and
INAF-Osservatorio Astronomico di Bologna, via Ranzani 1, 40127 Bologna, Italy\label{OABO}
\and
School of Physics, Astronomy \& Maths, University of Hertfordshire, Hatfield, Herts. AL10 9AB, UK\label{HERTS}
\and
Thueringer Landessternwarte, Sternwarte 5, 07778, Tautenburg, Germany \label{Tautenburg}
\and
CASU, Institute of Astronomy, Madingley Road, University of Cambridge, CB3 0HA, UK \label{CASU}
\and
IGAM, Institute of Physics, University of Graz, Universit\"{a}tsplatz. 5, A-8010 Graz, Austria \label{IGAM}
\and
Departament d’Astronomia i Meteorologia, Universitat de Barcelona, Avda. Diagonal 647, E-08028 Barcelona, Spain\label{AMB}
\and
Royal Observatory of Belgium, Ringlaan 3, 1180 Brussel, Belgium\label{BRUSSEL}
\and
Instituto de Astrof\'isica de Andaluc\'ia, CSIC, Apdo 3004, 18080 Granada, Spain\label{IAA}
\and
European Southern Observatory, Karl-Schwarzschild-Str. 2, 85748 Garching bei M\"{u}nchen, Germany\label{GARCHING}
\and
Department for Astrophysics, Nicolaus Copernicus Astronomical Center, ul. Rabia\'{n}ska 8, 87-100 Toru\'n, Poland\label{TORUN}
\and
INAF-Osservatorio Astrofisico di Arcetri, Largo E. Fermi, 5, 50125 Firenze, Italy\label{ARCETRI}
\and
Dipartimento di Fisica e Astronomia, Universit\`a di Bologna, via Ranzani 1, 40127 Bologna, Italy\label{UNIBO}
\and
Department of Astronomy, Indiana University, Bloomington, IN 47405, USA\label{INDIANA}
\and 
Massachusetts Institute of Technology and Kavli Institute for Astrophysics and Space Research, 77 Massachusetts Avenue, Cambridge, MA 02139, USA\label{MIT}
\and
Institute of Theoretical Physics and Astronomy, Vilnius University, Go\v{s}tauto 12, LT-01108 Vilnius, Lithuania\label{VILNIUS}
\and
INAF - Astrophysical Observatory of Torino, via Osservatorio 20, 10025 Pino Torinese, Italy\label{OATO}
\and
Section of Astrophysics, Astronomy and Mechanics, Department of Physics, University of Athens, GR-15784 Athens, Greece\label{ATHENS}
\and
INAF - Osservatorio Astrofisico di Catania, via S. Sofia 78, 95123 Catania, Italy\label{CATANIA}
\and
Institute of Astronomy, University of Cambridge, Madingley Road, Cambridge CB3 0HA, UK\label{IOA}
\and
Lund Observatory, Department of Astronomy and Theoretical Physics, Box 43, SE-221 00 \label{LUND}
\and
INAF - Osservatorio Astronomico di Palermo, Piazza del Parlamento 1, 90134, Palermo, Italy\label{PALERMO}
\and
Departamento de Astronom\'{i}a, Universidad de Concepci\'{o}n, Casilla 160-C, Concepci\'{o}n, Chile\label{CONCEPCION}
\and
Laboratoire Lagrange (UMR7293), Universit\'{e} de Nice Sophia Antipolis, CNRS, Observatoire de la C\^{o}te d'Azur, BP 4229, F-06304 Nice cedex 4, France\label{NICE}
\and
Spanish Virtual Observatory, Centro de Astrobiolog\'{i}a (INTA-CSIC), P.O. Box 78, 28691 Villanueva de la Ca\~nada, Madrid, Spain \label{SVO}
\and
Millennium Institute of Astrophysics, Av. Vicu\~{n}a Mackenna 4860, 782-0436 Macul, Santiago, Chile\label{MAS}
\and
Pontificia Universidad Cat\'{o}lica de Chile, Av. Vicu\~{n}a Mackenna 4860, 782-0436 Macul, Santiago, Chile\label{PUC}
}

\date{Received date / Accepted date }

\abstract
{Chemically inhomogeneous populations are observed in most globular clusters, but not in open clusters.
Cluster mass seems to play a key role in the existence of multiple populations.
}
{Studying the chemical homogeneity of the most massive open clusters is necessary to better understand  the mechanism of their formation and determine the mass limit under which clusters cannot host multiple populations. Here we studied NGC\,6705, that is a young and massive open cluster located towards the inner region of the Milky Way. This cluster is located inside the solar circle. This makes it an important tracer of the inner disk abundance gradient. }
{This study makes use of $BVI$ and $ri$ photometry and comparisons with theoretical isochrones to derive the age of NGC\,6705. We study the density profile of the cluster and the mass function to infer the cluster mass.
Based on abundances of the chemical elements distributed in the first internal data release of the Gaia-ESO Survey, we study elemental ratios and the chemical homogeneity of the red clump stars. Radial velocities enable us to study the rotation and internal kinematics of the cluster.}
{The estimated ages range from 250 to 316\,Myr, depending on the adopted stellar model. Luminosity profiles and mass functions show strong signs of mass segregation. We derive the mass of the cluster from its luminosity function and from the kinematics, finding values between 3700\,M$_{\odot}$ and 11\,000\,M$_{\odot}$. After selecting the cluster members from their radial velocities, we obtain a metallicity of [Fe/H]=0.10$\pm$0.06 based on 21 candidate members. Moreover, NGC\,6705 shows no sign of the typical correlations or anti-correlations between Al, Mg, Si, and Na, that are expected in multiple populations. This is consistent with our cluster mass estimate, which is lower than the required mass limit proposed in literature to develop multiple populations.}
{}


\maketitle{}

\section{Introduction}

Once thought to be the best example of simple stellar populations 
(coeval, mono-metallic systems), the globular clusters (GCs) have been shown
 to be instead complex objects. In particular, evidence that they host distinct populations
 (possibly, distinct generations) of stars has been mounting. 
\citet{Gratton04,Gratton12} present reviews based on spectroscopy, while
\citet{Piotto09} and \citet{Milone12} can be consulted for 
results based on photometry. 

While variations in iron content seem limited to very few cases,
 the most notable being $\omega$~Cen \citep[e.g.][]{Lee99,Johnson10} 
and M~22 \citep[e.g.][]{Marino11},
 GC stars show large star-to-star variations in some light elements,
 like C, N, O, Na, Mg, and Al. It has been known for a
long time that there are variations in the CN and CH bands strengths,
 with spreads and bimodal distributions \citep[e.g.][for reviews]{Kraft79,Kraft94},
and not all the variations can be attributed to internal processes like mixing.
Na and O have been extensively studied
as summarised in the review by \citet{Gratton04} and \citet{Carretta09a,Carretta09b},
 which show an anti-correlation: in the same GC there are stars with ``normal'' O and Na 
(normal with respect to the cluster metallicity) and stars showing a (strong) 
depletion in O and an enhancement in Na. Most of the stars have a modified composition.
A similar anti-correlation can be found between Mg and Al, but not 
in all clusters, and with lower depletions than in O \citep[e.g.][]{Carretta09b,Marino08}. 
Most importantly, these star-to-star variations occur also in unevolved, main-sequence stars,
 as demonstrated first by \citet{Gratton01} and \citet{Ramirez02} and later confirmed on
 much larger samples \citep[e.g.][]{Lind09,DOrazi10}. These low mass stars cannot have produced these variations, since their cores do not reach the high temperatures 
necessary for the relevant proton-capture reactions and, in any case, the stars lack the mixing
 mechanism needed to bring processed material to the surface.
This implies that the chemical inhomogeneities were already
present in the gas out of which these stars formed. This has led to the belief that GCs are made of at least two generations
of stars, with the second generation formed from gas polluted by the material processed by the first.
Exactly which kind of first-generation stars were the polluters is debated; 
the most promising candidates are intermediate mass asymptotic giant
 branch stars \citep{Ventura01} and fast rotating massive stars \citep{Decressin07}, 
but refinements and comparisons to robust observational constraints are required.

The multiple populations scenario for GCs requires 
that the cluster is massive enough to retain some of its primordial gas at the any of any subsequent star formation; indeed,
 existing models postulate that GCs were many times more massive at the time of 
their formations than they appear today \citep[e.g.][]{DErcole08}. Such models suffer from several drawkbacks. For instance, they do not explain the observed properties of the young massive clusters \citep{Bastian13}, or the abundance patterns of the halo stars \citep[see e.g.][]{Martell11}.
Recently, a mechanism based on accretion on circumstellar discs, not requiring multiple generation of stars has been presented by \citet{Bastian13b}. This model does not require that clusters were initially extremely massive. In any case, the efficiency of the accretion process depends on the density and velocity dispersion of the cluster, that both scale with the cluster mass, suggesting that more massive clusters should exhibit broader chemical dishonomegeneities.

\citet{Carretta10}, combining the data on 18 GCs of the FLAMES survey \citep{Carretta09a}
 and literature studies, showed that all Galactic GCs for which Na and O abundances were available presented a Na-O anti correlation, i.e., multiple populations. 
The only possible exceptions were Terzan\,7 and Palomar\,12, two young and low-mass GCs
 belonging to the Sagittarius dwarf galaxy. This apparent universality of the Na-O anti
 correlation was suggested to be a defining property of GCs, 
separating them from the other clusters where only a single population is present. 
Cluster mass seems to play a determinant role in this separation 
and, studies of low-mass GCs and massive and old open clusters (OCs) are needed to
 determine the mass threshold under which no multiple populations can be formed.
 This scenario needs however to be verified observationally,
which motivated further studies in this "grey zone" between the two kind of clusters. 
Recent additions include the GCs Rup\,106 \citep{Villanova13},  which shows 
no Na-O anti correlation, and Terzan\,8 \citep{Carretta13}, in which the second generation,
 if present, represents a minority, and two of the most massive
and oldest OCs, Berkeley\,39 and NGC\,6791.  High-resolution spectroscopy 
in 30 stars of Berkeley\,39 by \citet{Bragaglia12} showed a single, homogeneous population. 
\citet{Geisler12} have found signs of a bi-modal Na distribution in NGC\,6791,
 which would make it the first known OC with multiple populations. 
This suggestion has not been confirmed by another study \citep[][submitted]{Bragaglia13}
 where no variations exceeding the errors were found in the same cluster.
 Further studies are called for to settle the issue.

The Gaia-ESO Survey \citep[hereafter GES,][]{Gilmore12,Randich13} is a
large, public spectroscopic survey using the high-resolution
multi-object spectrograph FLAMES on the Very Large Telescope (ESO, Chile). It targets more than $10^5$
stars, covering the bulge, thick and thin discs, and halo components, and a
sample of about 100 open clusters (OCs) of all ages, metallicities, locations, and masses. The focus of the GES is to quantify the
kinematical and chemical element abundance distributions in the
different components of the Milky Way, thus providing a complement to the limited spectroscopic capabilities of the astrometric Gaia mission.

Here we focus on 
\object{NGC\,6705} (M\,11, Mel\,213, Cr\,391, OCl\,76),  a massive and concentrated
 OC located in the first galactic 
quadrant \citep{Messina10,Santos05} at the galactic coordinates $(l=27.3$, $b = -2.8)$.
This cluster, one of the three intermediate-age clusters observed in the first GES internal data release, contains a total of several thousands of solar masses 
\citep{McNamara77,Santos05}, which places it near the limit between 
the most massive OCs and the least massive GCs \citep{Bragaglia12}.
 Being located towards the central parts of the Galaxy it is projected against a 
dense background, but its rich main sequence and populated red clump clearly stand out
 in a color-magnitude diagram.

 NGC\,6705 is situated in a clear area and suffers 
from relatively little extinction for an object 
at such a low galactic latitude. Moreover, due to its position inside the solar circle, it is an important tracer of the inner disk gradient.
In a companion paper, \citet{Magrini14} compare the abundance patterns of the three intermediate-age OCs of the first GES data release (NGC\,6705, Tr\,20 and NGC\,4815) for Fe, Si, Mg, Ca, and Ti. They showed that each cluster has a unique pattern, with in particular a high value of [Mg/Fe] in NGC\,6705. Moreover, the abundances of two of these clusters are consistent with their galactocentric radius, whereas the abundances of NGC\,6705 are more compatible with a formation between 4 and 6\,kpc from the Galactic Center despite its current galactocentric radius of 6.9\,kpc.
The present paper focuses on the  determination of the  structural parameters, age and chemical abundances
of NGC\,6705, using VPHAS+, ESO 2.2.m telescope WFI photometry, and spectroscopic data from the Gaia-ESO Survey.

\begin{figure}[ht]
\begin{center}
\resizebox{\hsize}{!}{\includegraphics[scale=0.45]{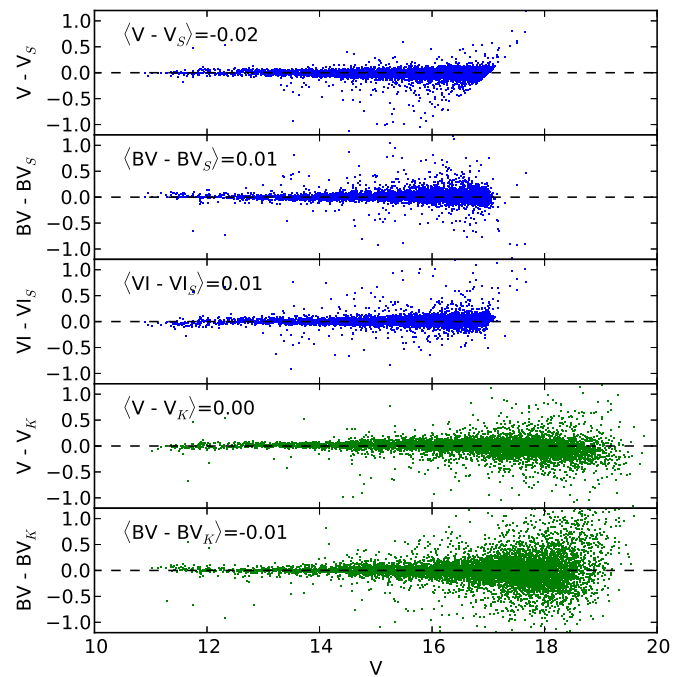}} \caption{\label{fig:comparePhot}\textit{Upper three panels:} comparison of our photometry with that of \citet{Sung99}. \textit{Lower two panels:} comparison with \citet{Koo07}. The labels indicate the residual zero point differences.}
\end{center}
\end{figure}

The photometric and spectroscopic  data are presented in Sect.~\ref{sec:data}. 
The membership determination is given in Sect.~\ref{sec:membership}. In Sect.~\ref{sec:extmap}
 we derive the extintion map in the region surrounding the cluster. In Sect.~\ref{sec:structure} we study the luminosity profile and mass segregation of the cluster, and provide cluster mass estimates. In Sect.~\ref{sec:kinematics} we look at the rotation and kinematics in the cluster. In Sect.~\ref{sec:params} we make use of isochrones to determine the distance and age of the cluster. Finally,
 the chemical abundances of Al, Mg, Si, and Na for NGC\,6705 are presented in Sect.~\ref{sec:homogeneity}.

\section{The data} \label{sec:data}

\subsection{WFI $BVI$ photometry}
The photometry used in this paper has been extracted from
archival images in $B$, $V$ and $I$ taken on the 13th of May 1999
with the Wide Field Imager at the MPG/ESO 2.2m Telescope for
the ESO Imaging Survey (ESO programme 163.O-0741(C), PI Renzini).
The sample of images comprise for each band two long exposure time
images (240~sec each for a total of 480~sec), in order to reach a
photometric precision of 20\% at V$\simeq21$ mag, and single short (20~sec)
exposure carefully chosen not to saturate the brighter targets.

Data reduction has been performed using the ESO/Alambic software
\citep{Vandame02} especially designed for mosaic CCD cameras
and providing fully astrometrically and photometrically calibrated
images with special care to the so-called ``illumination
correction''. The photometry was performed using the Daophot/Allstar 
software \citep{Stetson87} wrapped in an automatic procedure which performed
the PSF calculation and all the steps for extracting the final
magnitudes. The PSF photometry from the short and long exposures for each single
band have been combined in a single photometric dataset using the
Daomaster program.  The internal astrometric accuracy (both within the
short/long exposures and in the different bands) resulted to be better
than $0.05$~arcsec while the external comparison with the 2MASS
catalogue gave an rms of $\simeq0.20$~arcsec.

The observations consist in a single $33\arcmin \times 34\arcmin$ field centered on the cluster, covering completely the photometry of \citet{Sung99}. The geometry of the WFI imaging is shown in Fig~\ref{fig:extmapVPHAS}.
We tied our instrumental WFI photometry to their photometric calibration,
comparing the magnitudes of the common objects, calculating
solutions for the zero point and color term for each pass band. In the magnitude range V<14 the dispersion is of the order of 0.05 magnitudes, increasing to 0.1 for stars as faint  as V$\sim 17$ (see the upper panels of Fig.~\ref{fig:comparePhot}). 
Unfortunately, the $B$-band filter of WFI is quite far from the Johnson $B$ filter
with a large color term of $\simeq$-0.33\,mag. After applying the calibration, 
the $B$ magnitudes present a residual second-order trend in color which appears as a residual offset of -0.05\,mag for the extreme blue and red stars. We decided not to apply any 
correction for this effect, and instead we used the photometry of \citet{Sung99} for all objects with $V$ < 12 and $(B-V)$ > 1. For fainter stars, the photometries show good agreement, with a residual offset of $\simeq$0.02\,mag at $V$=14, increasing to 0.1\,mag at $V\simeq$17. The comparison between our photometry and that of \citet{Sung99} is shown in the upper three panels of Fig.~\ref{fig:comparePhot}.

Our data were also compared with the $BV$ photometric data of \citet{Koo07}, acquired in a search for variable stars. The comparison (lower panels of Fig.~\ref{fig:comparePhot}) also shows good agreement with no systematics, with a dispersion of 0.08 for $V<14$, increasing for fainter magnitudes.

\begin{figure}[ht]
\begin{center} \includegraphics[scale=0.55]{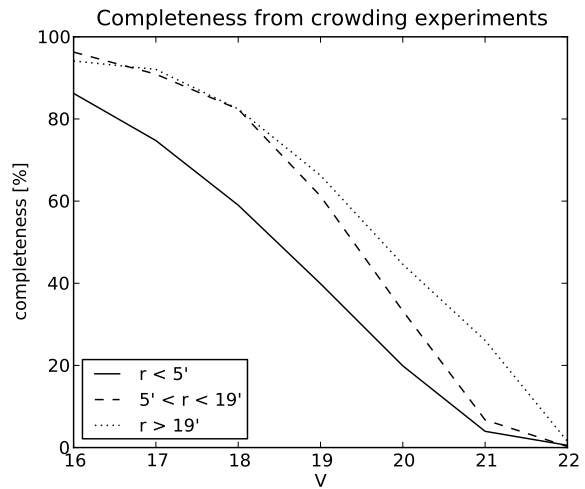} \caption{\label{completeness}Completeness of our photometry obtained from crowding experiments for different regions of the cluster.} \end{center}
\end{figure}

Since the cluster is very dense, our photometry suffers from crowding effects, especially in the central parts. The completeness of the photometry was estimated  as usual by randomly adding stars of magnitude $V$ ranging from 16 to 22 and running the source detection step again, then counting how many of the added stars were recovered. This method allows to estimate the completeness for each magnitude bin in a given region of the cluster. The completeness is better than 50\% up to magnitude $V\simeq18$ in the central regions and $V\simeq19.7$ in the less crowded outskirts (see Fig.~\ref{completeness}).

\begin{figure*}[ht]
\begin{center} \includegraphics[scale=0.7]{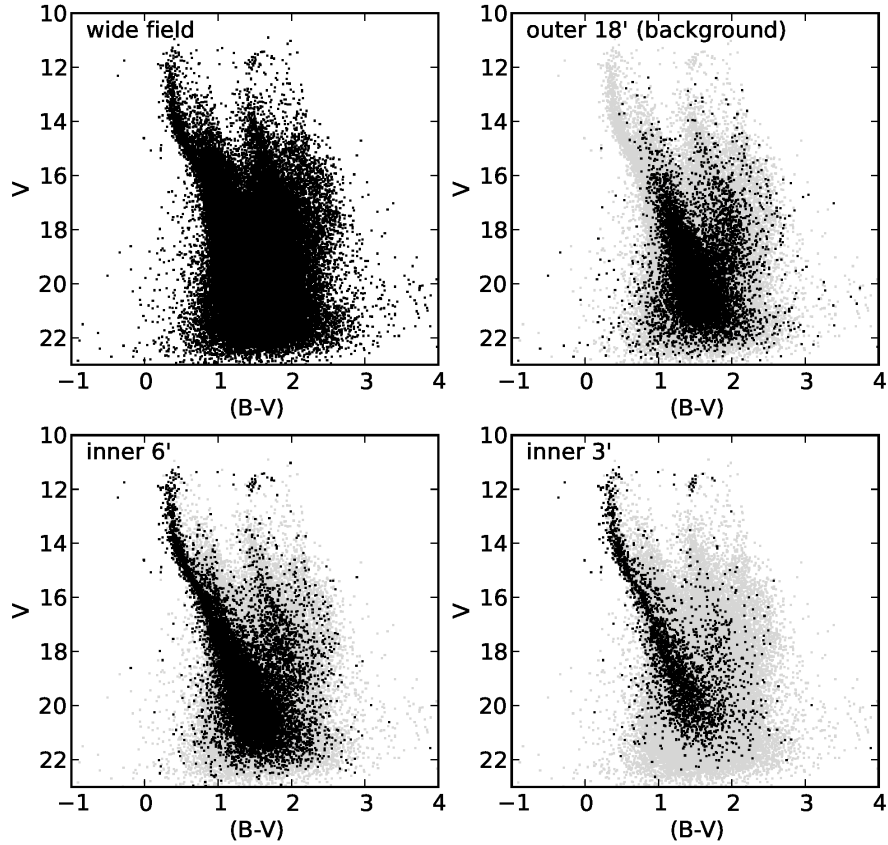} \caption{\label{cmd1}\textit{Top-left:} $BV$ CMD of all 123\,037 in our sample. \textit{Top-right:} CMD of the outer $18\arcmin$, which we consider as our background field. \textit{Bottom-left} and \textit{bottom-right:} CMD of the inner 6$\arcmin$ and $3\arcmin$. In the latter three panels, the grey points correspond to the wide-field CMD.} \end{center}
\end{figure*}

The data contain a total of 123\,037 stars. The $BV$ CMD shows a very clear main sequence standing out of the background as can be seen in Fig.~\ref{cmd1}. Evolved stars are also visible, with a red clump located around $(B-V,V)=(1.5,12)$. The lower panels of Fig.~\ref{cmd1} show that the main sequence can be easily followed down to $V$=16 in the inner regions of the cluster, while the top-right panel shows that the signature of the cluster is not visible outside of 18$\arcmin$.

\subsection{VPHAS+ $ri$ photometry}
The VPHAS+ observations used here are $r$- and $i$-band data from
three fields (numbered 212, 213, and 237) that overlap near the sky
position of NGC\,6705.  A general description of this public
survey continuing to execute using the OmegaCAM instrument on the
VLT Survey Telescope is provided by \citet{Drew14}.  Each pointing
captures a square degree at a time, with the center of NGC\,6705 falling 
within a couple of arcminutes of the southern edge of field
212, and towards the boundary with 237. The data were all obtained at a
time of very good seeing (0.5 to 0.6 arcseconds) during the night of
7th July 2012 when the moon was relatively bright (FLI = 0.77). 
In $r$, the
formal 10-$\sigma$ magnitude limit falls between 20.5 and 20.8 with a
likely completeness limit of $\sim$19.5, while these limits are
approximately 0.5--0.7 magnitudes brighter in the $i$ band. All magnitudes are specified
in the Vega system.

\begin{figure}[ht]
\begin{center} \resizebox{\hsize}{!}{\includegraphics[scale=0.6]{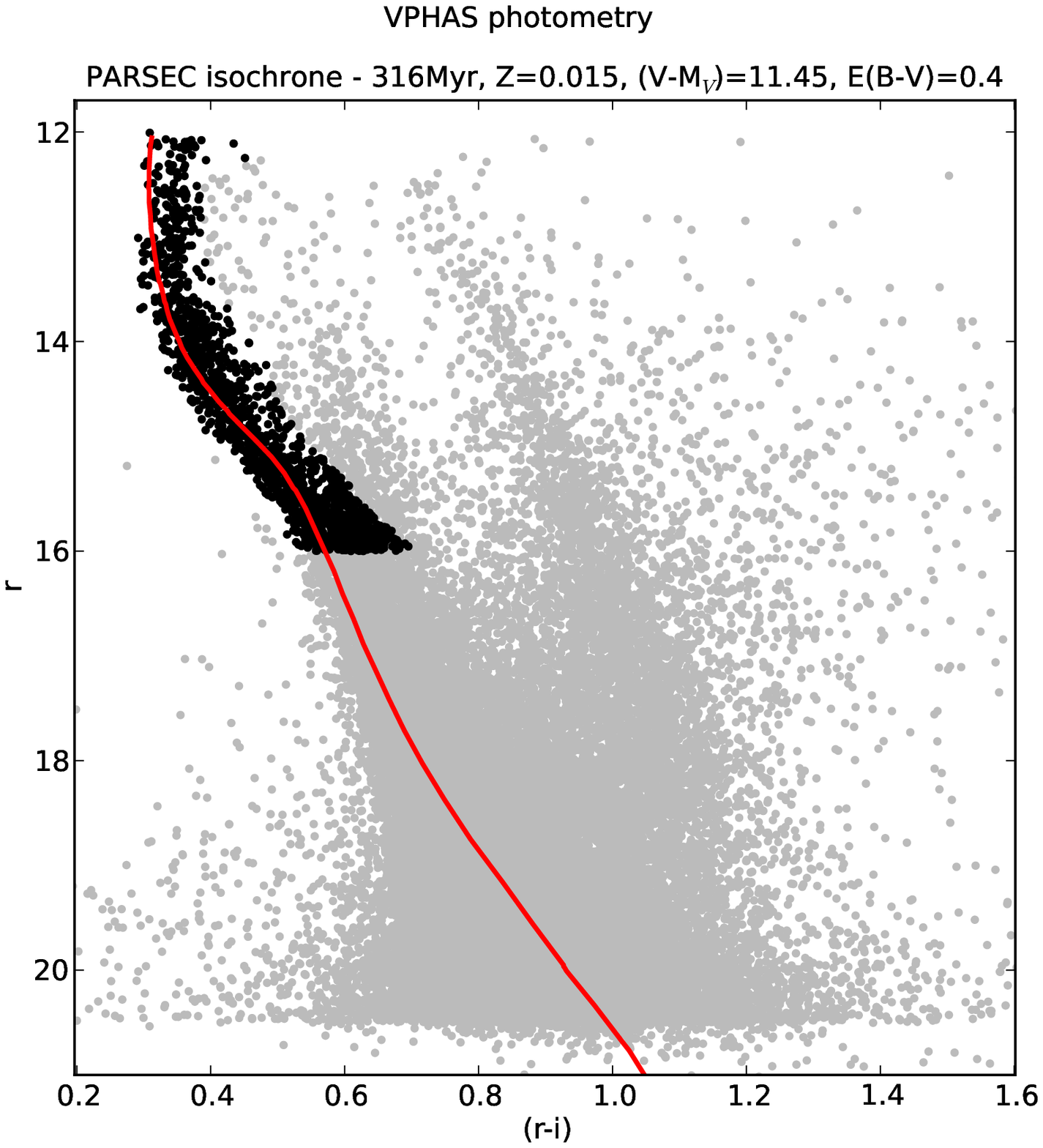}} \caption{\label{fig:VPHASCMD}VPHAS+ $ri$ CMD of the central 12$\arcmin$ of the cluster. The spatial distribution of the selected stars (black points) is shown in Fig.~\ref{fig:extmapVPHAS}. The solide line corresponds to the PARSEC isochrone of best parameters, shifted for distance modulus and extinction (see Sect.~\ref{sec:params} for detail). Stars brigther than $r=12$ suffer from saturation and are not shown here.} \end{center}
\end{figure}

The CMD of the inner 12$\arcmin$ is shown in Fig.~\ref{fig:VPHASCMD}. The data are used in Sect.~\ref{sec:extmap} to derive the extinction map of the region surrounding the cluster.
Due to the very good seeing, stars brighter than $r \sim 12$ are partially saturated in both $r$ and $i$ bands, making the position of the main sequence turn-off point and red clump stars quite uncertain in the CMD.

\subsection{GES spectroscopic data}
The GES makes use of the GIRAFFE (R$\sim$20\,000) and UVES (R$\sim$47\,000) 
spectrographs of the VLT UT-2. 
The GES consortium is structured in several working groups (WGs). 
The reduction of the GIRAFFE and UVES 
spectra by WG7 is described in Lewis et al. (in prep.)
 and \citet{Sacco14}, respectively. 
 The  analysis is performed independently by teams using different methods, but make use of the same model atmospheres \citep[MARCS,][]{MARCS} and the same line list \citep{Heiter14}.
 The results are then gathered and controlled to produce a 
homogenised set \citep[for a comparison of the results obtained with different spectroscopic methods, see for instance][]{Jofre13}. The analysis of the GIRAFFE data by WG10 is described 
in Recio-Blanco et al. (in prep.), 
and the  analysis of UVES spectra by WG11 is described in Smiljanic et al. (in prep.).

Stellar parameters and elemental abundances for the stars observed 
during the first six months of the campaign were delivered to the members
 of the collaboration as an internal data release (internal data release 1, or GESviDR1Final)
 for the purpose of science verification and validation.

The target selection for NGC\,6705  and the exposure times for the various setups are described in Bragaglia et al. (in prep.). For the GIRAFFE targets, potential members were selected on the basis of their optical and infrared photometry following the cluster main sequence, and the proper motions from the UCAC4 catalog \citep{UCAC4} were used in order to discard objects whose proper motions are more than five sigma from the cluster centroid.
In total, 1028 main-sequence stars were observed with 8 GIRAFFE setups. 
All the red giants located in the clump region in the inner 12$\arcmin$ were observed with the UVES setup 580 (25 targets). The number of targets observed with each GIRAFFE setup is summarised in Table~\ref{tab:targets}. The coordinates, $B$, $V$, $J$, $H$, and $K$ magnitudes, radial velocities and membership probabilities (see Sect.~\ref{sec:membership}) are shown in Table~\ref{tab:1028GIRAFFE}.

\begin{table}
\begin{center}
	\caption{ \label{tab:targets}Summary of GES GIRAFFE observations}
	\small\addtolength{\tabcolsep}{-2pt}
	\begin{tabular}{ l c c c c c }
	\hline
	\hline
	Setup   & Central $\lambda$ & Spectral range        & R	  & nb Stars  & Median S/N\\
	        & [nm]          & [nm]          &   	  & 	&   \\
	\hline
	HR3	& 412.4		& 403.3--420.1	& 24\,800 & 166 & 24\\
	HR5A 	& 447.1 	& 434.0--458.7	& 18\,470 & 166	& 25\\ 	
	HR6 	& 465.6 	& 453.8--475.9	& 20\,350 & 166	& 21\\
	HR9B 	& 525.8 	& 514.3--535.6	& 25\,900 & 526 & 37\\
	HR10 	& 447.1 	& 434.0--458.7	& 18\,470 & 284	& 50\\
	HR14A 	& 651.5 	& 630.8--670.1	& 17\,740 & 166 & 44\\
	HR15N	& 665.0		& 647.0--679.0  & 17\,000 & 1028 & 52\\
	HR21 	& 875.7 	& 434.0--458.7	& 18\,470 & 284 & 55\\
	\hline
	\end{tabular}
\end{center}
\end{table}

\begin{table*}
\begin{center}
	\caption{ \label{tab:1028GIRAFFE}Positions, $BV$ photometry, radial velocities and membership probabilities of the GIRAFFE targets.}
	\small\addtolength{\tabcolsep}{-2pt}
	\begin{tabular}{l l l l l l c}
	\hline\hline
	  \multicolumn{1}{c}{Star} &
	  \multicolumn{1}{c}{RA} &
	  \multicolumn{1}{c}{DEC} &
	  \multicolumn{1}{c}{$B$} &
	  \multicolumn{1}{c}{$V$} &
	  \multicolumn{1}{c}{RV} &
	  \multicolumn{1}{c}{Membership probability} \\
	\hline
	  18505884-0614409 & 282.7452 & -6.2447 & 15.160 & 14.661 & 34.3 $\pm$ 1.3 & 0.91\\
	  18505976-0616255 & 282.7490 & -6.2738 & 14.855 & 14.358 & 31.8 $\pm$ 2.2 & 0.81\\
	  18505998-0617359 & 282.7499 & -6.2933 & 14.743 & 14.288 & 37.4 $\pm$ 0.9 & 0.92\\
	  \multicolumn{7}{c}{...} \\
	  18514946-0620231 & 282.9561 & -6.3398 & 19.767 & 18.445 & 120.7 $\pm$ 1.5 & 0.0\\
	\hline
	\end{tabular}
\tablefoot{The full table (containing 1028 rows) is available in the electronic version of this article.}
\end{center}
\end{table*}

\begin{figure*}[ht]
\begin{center} \includegraphics[scale=0.65]{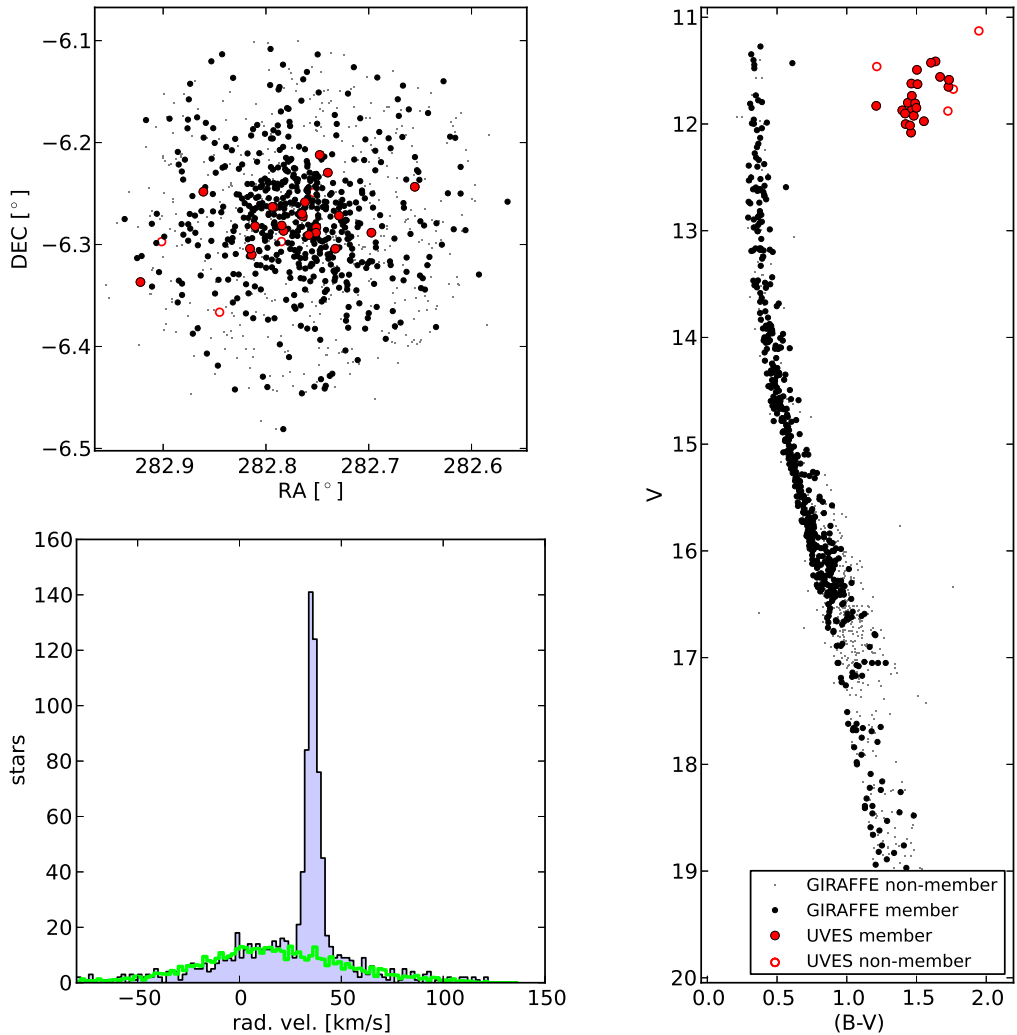} \caption{\label{fig:recapUvesGiraffe}\textit{Top-left}: positions of the GIRAFFE (black dots for members, grey points for non-members) and UVES targets (red dots for members, empy red symbols for non-members). \textit{Bottom-left:} radial velocity distribution of the GIRAFFE targets. The green line shows the expected radial velocity distribution of the field stars from the Besan\c{c}on model \citep{Robin03}. \textit{Right:} CMD of all the GES stars.} \end{center}
\end{figure*}

The spatial distribution of the targets is shown in the top-left panel of 
Fig.~\ref{fig:recapUvesGiraffe},
 and the $BV$ photometry of the targets is the WFI photometry presented in the present paper.
The bottom-left panel of Fig.~\ref{fig:recapUvesGiraffe} shows the radial velocity (RV) 
distribution of all the GIRAFFE targets. From the GESviDR1Final data release, we have RVs for 1028 GIRAFFE targets and 25 UVES targets.

The 25 UVES stars have well-defined values of effective temperature, 
surface gravity, and microturbulence derived directly from these spectra (summarised in Table~\ref{tab:summaryM11}).
While elemental abundances are available for various elements in the GESviDR1final data release, this work focused on the abundances of Fe, Al, Mg, Na, and Si (Table~\ref{tab:abundancesM11}). The reference solar abundances used by GES are those of \citet{Grevesse07}.

Due to the fact that the main-sequence targets are hot, fast-rotating stars, preliminary metallicity determinations from GIRAFFE are for the moment available for 255 GIRAFFE stars only in GESviDR1Final, with large quoted uncertainties.
These data will be re-analysed in subsequent GES data releases.
We have decided not to consider these results and focus only on the metallicity determination obtained from the UVES spectra.

\begin{table*}
\begin{center}
	\caption{ \label{tab:summaryM11}Photometry and stellar parameters of the UVES stars.}
	\small\addtolength{\tabcolsep}{-2pt}
	\begin{tabular}{ l l l l l l l l c l l l l }
	\hline
	\hline
	\multicolumn{1}{c}{Star} & \multicolumn{1}{c}{RA} & \multicolumn{1}{c}{DEC} & \multicolumn{1}{c}{$B$} & \multicolumn{1}{c}{$V$} & \multicolumn{1}{c}{$J$} & \multicolumn{1}{c}{$H$} & \multicolumn{1}{c}{$K_{s}$} & \multicolumn{1}{c}{RV} & \multicolumn{1}{c}{T$_{\textrm{eff}}$} & \multicolumn{1}{c}{log $g$} & \multicolumn{1}{c}{[Fe/H]} & \multicolumn{1}{c}{v$_{mic}$}\\
	 & \multicolumn{1}{c}{[\degr]} & \multicolumn{1}{c}{[\degr]} & & &  &   &   & \multicolumn{1}{c}{[km\,s$^{-1}$]} & \multicolumn{1}{c}{[K]} & & & \multicolumn{1}{c}{[km\,s$^{-1}$]}\\
	\hline
	\multicolumn{13}{c}{Members} \\
	\hline
	18503724-0614364 & 282.6552 & -6.2434 & 13.423 & 12.001 & 9.427 & 8.808 & 8.616 & 35.2 & 4820 $\pm$ 71 & 2.42 $\pm$ 0.21 & 0.03 $\pm$ 0.14 & 1.82 $\pm$ 0.13 \\
	18504737-0617184 & 282.6974 & -6.2884 & 13.381 & 11.652 & 8.527 & 7.808 & 7.523 & 31.8 & 4325 $\pm$ 130 & 1.72 $\pm$ 0.29 & 0.03 $\pm$ 0.15 & 1.56 $\pm$ 0.16 \\
	18505494-0616182 & 282.7289 & -6.2717 & 13.328 & 11.860 & 9.199 & 8.498 & 8.318 & 34.8 & 4689 $\pm$ 109 & 2.37 $\pm$ 0.43 & 0.13 $\pm$ 0.09 & 1.46 $\pm$ 0.12 \\
	18505581-0618148 & 282.7325 & -6.3041 & 13.050 & 11.414 & 8.557 & 7.895 & 7.698 & 35.1 & 4577 $\pm$ 139 & 2.23 $\pm$ 0.31 & 0.17 $\pm$ 0.18 & 1.60 $\pm$ 0.24 \\
	18505755-0613461 & 282.7398 & -6.2295 & 13.041 & 11.830 & 9.426 & 8.852 & 8.675 & 30.4 & 4873 $\pm$ 114 & 2.37 $\pm$ 0.32 & 0.03 $\pm$ 0.14 & 1.33 $\pm$ 0.19 \\
	18505944-0612435 & 282.7477 & -6.2121 & 13.270 & 11.872 & 9.330 & 8.722 & 8.523 & 34.6 & 4925 $\pm$ 177 & 2.56 $\pm$ 0.39 & 0.19 $\pm$ 0.18 & 1.50 $\pm$ 0.50 \\
	18510023-0616594 & 282.7510 & -6.2832 & 13.319 & 11.586 & 8.524 & 7.776 & 7.589 & 35.2 & 4433 $\pm$ 95 & 1.94 $\pm$ 0.47 & 0.17 $\pm$ 0.12 & 1.50 $\pm$ 0.14 \\
	18510032-0617183 & 282.7513 & -6.2884 & 13.542 & 12.081 & 9.368 & 8.751 & 8.549 & 35.2 & 4850 $\pm$ 100 & 2.38 $\pm$ 0.21 & 0.07 $\pm$ 0.15 & 1.60 $\pm$ 0.33 \\
	18510200-0617265 & 282.7583 & -6.2907 & 13.030 & 11.426 & 8.446 & 7.766 & 7.493 & 32.0 & 4415 $\pm$ 87 & 2.35 $\pm$ 0.45 & 0.18 $\pm$ 0.14 & 1.48 $\pm$ 0.07 \\
	18510289-0615301 & 282.7620 & -6.2584 & 13.467 & 12.014 & 9.389 & 8.758 & 8.543 & 32.8 & 4750 $\pm$ 112 & 2.40 $\pm$ 0.28 & 0.05 $\pm$ 0.07 & 1.45 $\pm$ 0.13 \\
	18510341-0616202 & 282.7642 & -6.2723 & 13.239 & 11.801 & 9.216 & 8.579 & 8.386 & 36.3 & 4975 $\pm$ 146 & 2.50 $\pm$ 0.30 & 0.07 $\pm$ 0.15 & 1.94 $\pm$ 0.27 \\
	18510358-0616112 & 282.7649 & -6.2698 & 13.320 & 11.902 & 9.357 & 8.691 & 8.511 & 34.7 & 4832 $\pm$ 79 & 2.31 $\pm$ 0.31 & 0.15 $\pm$ 0.08 & 1.62 $\pm$ 0.19 \\
	18510786-0617119 & 282.7828 & -6.2866 & 13.084 & 11.621 & 9.030 & 8.399 & 8.206 & 33.6 & 4768 $\pm$ 53 & 2.11 $\pm$ 0.19 & 0.03 $\pm$ 0.14 & 1.80 $\pm$ 0.28 \\
	18510833-0616532 & 282.7847 & -6.2814 & 13.202 & 11.736 & 8.911 & 8.311 & 8.227 & 33.3 & 4750 $\pm$ 112 & 2.25 $\pm$ 0.22 & 0.18 $\pm$ 0.10 & 1.60 $\pm$ 0.25 \\
	18511013-0615486 & 282.7922 & -6.2635 & 12.996 & 11.493 & 8.613 & 7.933 & 7.705 & 36.9 & 4439 $\pm$ 59 & 1.87 $\pm$ 0.53 & 0.10 $\pm$ 0.12 & 1.50 $\pm$ 0.10 \\
	18511048-0615470 & 282.7937 & -6.2631 & 13.134 & 11.627 & 8.817 & 8.224 & 7.991 & 33.3 & 4744 $\pm$ 122 & 2.12 $\pm$ 0.33 & 0.05 $\pm$ 0.14 & 1.70 $\pm$ 0.30 \\
	18511452-0616551 & 282.8105 & -6.2820 & 13.403 & 11.923 & 9.263 & 8.620 & 8.420 & 35.1 & 4800 $\pm$ 59 & 2.40 $\pm$ 0.25 & 0.11 $\pm$ 0.08 & 1.69 $\pm$ 0.20 \\
	18511534-0618359 & 282.8139 & -6.3100 & 13.527 & 11.974 & 9.163 & 8.507 & 8.303 & 33.7 & 4755 $\pm$ 57 & 2.16 $\pm$ 0.21 & 0.05 $\pm$ 0.22 & 1.79 $\pm$ 0.17 \\
	18511571-0618146 & 282.8155 & -6.3041 & 13.297 & 11.807 & 9.088 & 8.445 & 8.211 & 35.0 & 4710 $\pm$ 159 & 2.27 $\pm$ 0.30 & 0.15 $\pm$ 0.12 & 1.60 $\pm$ 0.18 \\
	18512662-0614537 & 282.8609 & -6.2482 & 13.228 & 11.559 & 8.611 & 7.921 & 7.700 & 33.8 & 4459 $\pm$ 91 & 2.10 $\pm$ 0.48 & 0.12 $\pm$ 0.17 & 1.48 $\pm$ 0.19 \\
	18514130-0620125 & 282.9221 & -6.3368 & 13.348 & 11.849 & 9.185 & 8.570 & 8.361 & 33.2 & 4671 $\pm$ 140 & 2.20 $\pm$ 0.28 & 0.07 $\pm$ 0.19 & 1.62 $\pm$ 0.20 \\
	\hline
	\multicolumn{13}{c}{Non-members} \\
	\hline
	18510093-0614564 & 282.7539 & -6.2490 & 12.677 & 11.462 & 9.016 & 8.395 & 8.205 & 41.4 & 4755 $\pm$ 35 & 2.30 $\pm$ 0.24 & -0.10 $\pm$ 0.10 & 1.50 $\pm$ 0.15 \\
	18510837-0617495 & 282.7849 & -6.2971 & 13.438 & 11.674 & 8.469 & 7.649 & 7.429 & -72.5 & 4217 $\pm$ 83 & 1.62 $\pm$ 0.33 & -0.10 $\pm$ 0.12 & 1.47 $\pm$ 0.16 \\
	18512283-0621589 & 282.8451 & -6.3664 & 13.604 & 11.879 & 8.831 & 8.084 & 7.872 & 3.4 & 4305 $\pm$ 95 & 2.08 $\pm$ 0.34 & 0.18 $\pm$ 0.18 & 1.61 $\pm$ 0.21 \\
	18513636-0617499 & 282.9015 & -6.2972 & 13.075 & 11.128 & 7.295 & 6.451 & 6.127 & -1.4 & 4041 $\pm$ 225 & 1.57 $\pm$ 0.42 & -0.20 $\pm$ 0.20 & 1.45 $\pm$ 0.03 \\
	\hline
	\end{tabular}
\tablefoot{The photometry was not corrected for extinction. The nominal uncertainty on UVES radial velocities is 0.6\,km\,s$^{-1}$.}
\end{center}
\end{table*}

\begin{table*}
\begin{center}
	\caption{ \label{tab:abundancesM11}Elemental abundances for the UVES stars.}
	\small\addtolength{\tabcolsep}{-2pt}
	\begin{tabular}{ l l l l l l}
	\hline
	\hline
	\multicolumn{1}{c}{Star} & \multicolumn{1}{c}{[Fe/H]} & \multicolumn{1}{c}{[Al/Fe]} & \multicolumn{1}{c}{[Mg/Fe]} & \multicolumn{1}{c}{[Na/Fe]} & \multicolumn{1}{c}{[Si/Fe]} \\
	\hline
	\multicolumn{6}{c}{Members} \\
	\hline
		18503724-0614364 & 0.03 $\pm$ 0.14 & 0.12 $\pm$ 0.15 & 0.23 $\pm$ 0.17 & 0.51 $\pm$ 0.17 & 0.09 $\pm$ 0.15 \\
		18504737-0617184 & 0.03 $\pm$ 0.15 & 0.17 $\pm$ 0.17 & 0.23 $\pm$ 0.15 & 0.74 $\pm$ 0.21 & 0.03 $\pm$ 0.17 \\
		18505494-0616182 & 0.13 $\pm$ 0.09 & 0.26 $\pm$ 0.09 & 0.14 $\pm$ 0.09 & 0.5 $\pm$ 0.11 & 0.07 $\pm$ 0.09 \\
		18505581-0618148 & 0.17 $\pm$ 0.18 & 0.10$\pm$ 0.18 & 0.12 $\pm$ 0.20 & 0.32 $\pm$ 0.21 & 0.05 $\pm$ 0.19 \\
		18505755-0613461 & 0.03 $\pm$ 0.14 & -0.04 $\pm$ 0.38 & 0.10$\pm$ 0.15 & 0.52 $\pm$ 0.14 & 0.07 $\pm$ 0.15 \\
		18505944-0612435 & 0.19 $\pm$ 0.18 & 0.14 $\pm$ 0.18 & 0.14 $\pm$ 0.22 & 0.49 $\pm$ 0.19 & -0.04 $\pm$ 0.19 \\
		18510023-0616594 & 0.17 $\pm$ 0.12 & 0.21 $\pm$ 0.12 & 0.20 $\pm$ 0.16 & 0.39 $\pm$ 0.13 & 0.02 $\pm$ 0.18 \\
		18510032-0617183 & 0.07 $\pm$ 0.15 & 0.20 $\pm$ 0.15 & 0.21 $\pm$ 0.18 & 0.53 $\pm$ 0.16 & 0.02 $\pm$ 0.16 \\
		18510200-0617265 & 0.18 $\pm$ 0.14 & 0.13 $\pm$ 0.15 & 0.16 $\pm$ 0.16 & 0.17 $\pm$ 0.14 & 0.02 $\pm$ 0.15 \\
		18510289-0615301 & 0.05 $\pm$ 0.07 & 0.22 $\pm$ 0.07 & 0.13 $\pm$ 0.13 & 0.53 $\pm$ 0.11 & 0.05 $\pm$ 0.09 \\
		18510341-0616202 & 0.07 $\pm$ 0.15 & 0.19 $\pm$ 0.15 & 0.30 $\pm$ 0.16 & 0.54 $\pm$ 0.17 & -0.07 $\pm$ 0.17 \\
		18510358-0616112 & 0.15 $\pm$ 0.08 & 0.18 $\pm$ 0.08 & 0.12 $\pm$ 0.10 & 0.49 $\pm$ 0.10 & 0.07 $\pm$ 0.11 \\
		18510786-0617119 & 0.03 $\pm$ 0.14 & 0.25 $\pm$ 0.14 & 0.42 $\pm$ 0.14 & 0.58 $\pm$ 0.18 & 0.08 $\pm$ 0.18 \\
		18510833-0616532 & 0.18 $\pm$ 0.10 & 0.18 $\pm$ 0.10 & 0.19 $\pm$ 0.14 & 0.51 $\pm$ 0.13 & -0.06 $\pm$ 0.12 \\
		18511013-0615486 & 0.10 $\pm$ 0.12 & 0.23 $\pm$ 0.12 & 0.23 $\pm$ 0.16 & 0.52 $\pm$ 0.15 & 0.07 $\pm$ 0.14 \\
		18511048-0615470 & 0.05 $\pm$ 0.14 & 0.23 $\pm$ 0.15 & 0.39 $\pm$ 0.14 & 0.54 $\pm$ 0.21 & -0.03 $\pm$ 0.23 \\
		18511452-0616551 & 0.11 $\pm$ 0.08 & 0.15 $\pm$ 0.11 & 0.15 $\pm$ 0.29 & 0.55 $\pm$ 0.16 & 0.06 $\pm$ 0.11 \\
		18511534-0618359 & 0.05 $\pm$ 0.22 & 0.25 $\pm$ 0.22 & 0.33 $\pm$ 0.24 & 0.49 $\pm$ 0.23 & 0.08 $\pm$ 0.24 \\
		18511571-0618146 & 0.15 $\pm$ 0.12 & 0.20 $\pm$ 0.13 & 0.09 $\pm$ 0.17 & 0.45 $\pm$ 0.14 & 0.0 $\pm$ 0.13 \\
		18512662-0614537 & 0.12 $\pm$ 0.17 & 0.29 $\pm$ 0.17 & 0.24 $\pm$ 0.22 & 0.59 $\pm$ 0.21 & 0.13 $\pm$ 0.22 \\
		18514130-0620125 & 0.07 $\pm$ 0.19 & 0.21 $\pm$ 0.19 & 0.11 $\pm$ 0.21 & 0.45 $\pm$ 0.20 & 0.01 $\pm$ 0.20 \\
	\hline
	$\mu$		         & 0.10 $\pm$ 0.04 & 0.20 $\pm$ 0.04 & 0.19 $\pm$ 0.05 & 0.48 $\pm$ 0.05 & 0.04 $\pm$ 0.04 \\ 
	$\sigma$	         & 0$^{+0.04}$     & 0$^{+0.05}$     & 0$^{+0.07}$     & 0$^{+0.06}$     & 0$^{+0.05}$     \\ 
	\hline
	\multicolumn{6}{c}{Non-members} \\
	\hline
		18510093-0614564 & -0.10$\pm$ 0.10 & 0.23 $\pm$ 0.11 & 0.11 $\pm$ 0.16 & 0.45 $\pm$ 0.13 & 0.07 $\pm$ 0.12 \\
		18510837-0617495 & -0.10$\pm$ 0.12 & 0.20 $\pm$ 0.12 & 0.28 $\pm$ 0.14 & 0.32 $\pm$ 0.12 & 0.02 $\pm$ 0.13 \\
		18512283-0621589 & 0.18 $\pm$ 0.18 & 0.27 $\pm$ 0.20 & 0.24 $\pm$ 0.20 & 0.38 $\pm$ 0.23 & 0.06 $\pm$ 0.20 \\
		18513636-0617499 & -0.20 $\pm$ 0.20 & 0.53 $\pm$ 0.21 & 0.43 $\pm$ 0.20 & 0.41 $\pm$ 0.23 & 0.06 $\pm$ 0.21 \\
	\hline
	\hline
	\multicolumn{1}{c}{Sun$^{a}$} & \multicolumn{1}{c}{7.45} & \multicolumn{1}{c}{6.37} & \multicolumn{1}{c}{7.53} & \multicolumn{1}{c}{6.17} & \multicolumn{1}{c}{7.51} \\
	\hline
	\end{tabular}
\tablefoot{$^{a}$The solar reference abundances are those of \citet{Grevesse07}. $\mu$ and $\sigma$ are the intrinsic mean and dispersion, respectively (cf. Sect.~\ref{sec:homogeneity}).}
\end{center}
\end{table*}

\subsection{HARPS Archive data} \label{sec:dataHARPS}

\begin{figure}[ht]
\begin{center} \resizebox{\hsize}{!}{\includegraphics[scale=0.57]{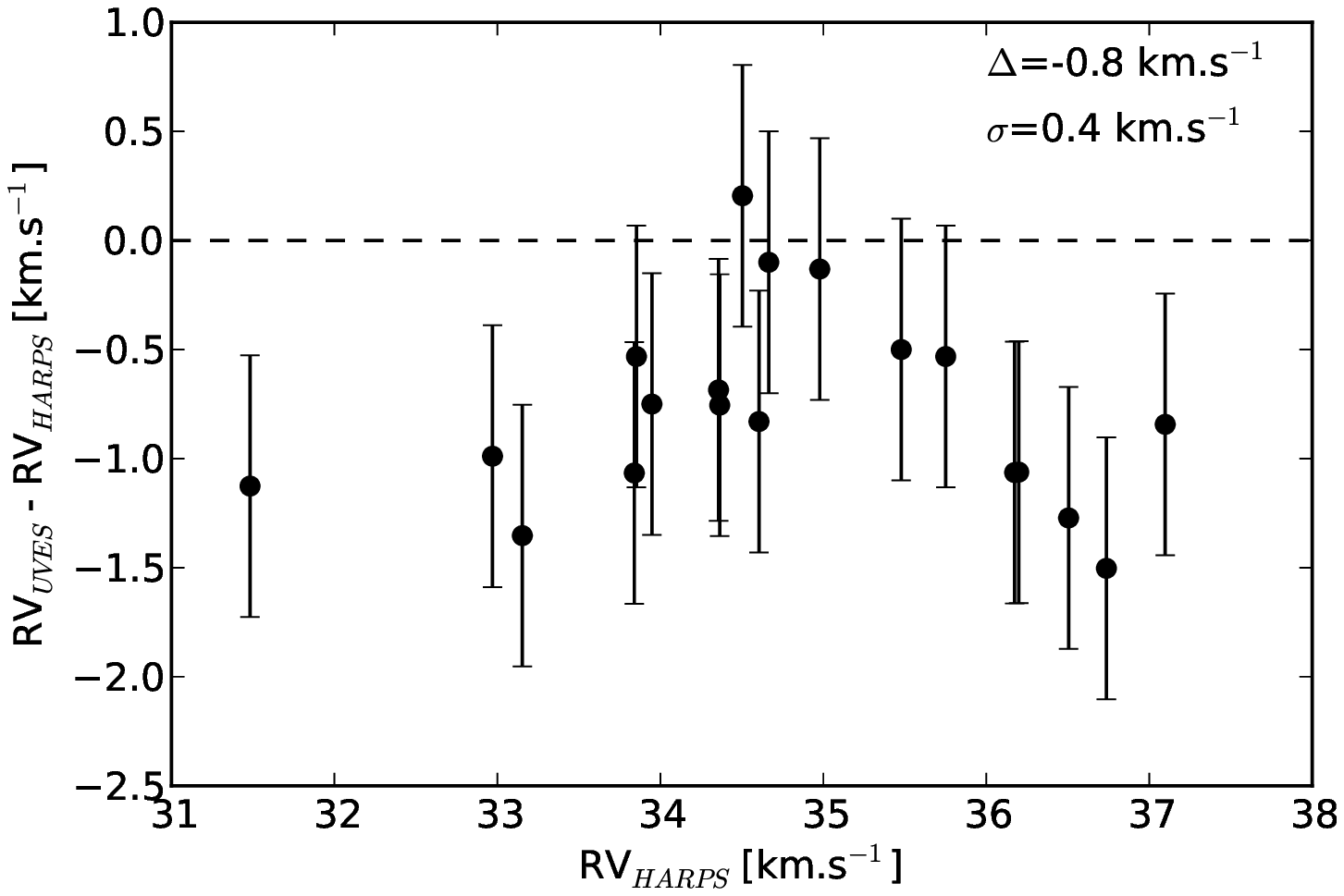}} \caption{\label{fig:radvelHARPSvsUVES}Comparison between the radial velocity measurements from HARPS spectra and UVES spectra for 19 red giants.} \end{center}
\end{figure}

We compared our results with those
obtained using an 
HARPS archive data set taken by the program Search for
Planets around Evolved Intermediate-Mass Stars \citep{Lovis07}.
 29 stars in
the cluster have been monitored for a total of more than 70 hours 
of on-target integrations
and 271 spectra taken. The observing program covered several periods 
from P79 (2007) to
P83 (2009).
In the ESO archive, the observed dataset is already reduced, providing a  measure of the RVs. 
Owing to its high resolving power (R$\sim$115\,000) and simultaneous wavelength calibration, HARPS
delivers a typical radial velocity error of about 0.06 km\,s$^{-1}$ for these stars.
19 of them stars were also observed by GES with the UVES instrument,
 allowing for a sanity check of our UVES measurements, but none were observed with GIRAFFE.
Fig.~\ref{fig:radvelHARPSvsUVES} shows a comparison between the RVs 
derived from UVES and HARPS spectra. 
The nominal uncertainty on the UVES RVs is 0.6\,km\,s$^{-1}$.
 The measurements from UVES are systematically lower by 0.8\,km\,s$^{-1}$ on average,
 with a standard deviation of 0.4\,km\,s$^{-1}$. We corrected for the differences in zero-point between the GIRAFFE, UVES and HARPS radial velocities when using them together later in this study.

\section{Membership} \label{sec:membership}

The spectroscopic targets were photometrically selected to be likely cluster members, but this selection obviously includes a significant number of field stars, that can be separated from the cluster stars on the basis of their RVs.
In Fig.~\ref{fig:recapUvesGiraffe}, a green line shows the RV distribution 
of the field stars expected from the Besan\c{c}on model \citep{Robin03}. 
We selected from the simulation the stars belonging to the bright main sequence region,
 with $B-V$ < 1.2 and $V$ < 17 
(the region of the CMD covered by the majority of the GIRAFFE targets) 
and scaled the distribution so that its tails 
(RV $<$ 20\,km\,s$^{-1}$ and RV $>$ 50\,km\,s$^{-1}$)
 contain the same number of objects as those of the observed distribution.
The Besan\c{c}on model reproduces very well the observed RV distribution of the background, while the signature of the cluster is clearly visible as a peak around 36\,km\,s$^{-1}$.

We performed a membership determination, for GIRAFFE and UVES stars independently.

\subsection{Membership of the GIRAFFE stars} \label{sec:membershipG}
We determined the membership of the GIRAFFE stars from the HR15N radial velocities only, because of the good signal-to-noise ratio (S/N) of these spectra and because all 1028 GIRAFFE stars were observed with this setup. The analysis of the data obtained with the other gratings will be available in further data releases.

We applied a classical parametric procedure where 
the radial velocity distribution is fitted with two Gaussian components: 
one for the cluster members and one for the field stars.
 We followed the procedure described in \citet{Cabrera85}, 
but based on the RVs only. 
The RVs of stars that were observed with 
different GIRAFFE gratings are the 2-sigma clipped averages. 
The method computes the membership probabilities through an iterative method.
 The probability density function (hereafter PDF) model is defined as:  

\begin{equation}
   \phi_i(v_i) = n_c \phi_{i,c}(v_i)+ n_f \phi_{i,f}(v_i)
\end{equation}

\noindent where \textit{$n_c$} and \textit{$n_f$} are the priors of the
 cluster member and field stars distributions respectively, 
\textit{$\phi_i(v_i)$} the PDF for the whole sample and, 
\textit{$\phi_{i,c}(v_i)$} and \textit{$\phi_{i,f}(v_i)$} 
the PDFs for the cluster members and the field stars related to the \textit{i}-th star.

The membership probabilities are obtained making use of 
these PDFs and by using Bayes' theorem as follows:

\begin{equation}
   \textit{P$_c(v_i)$} = \frac{n_c\phi_{i,c}(v_i)}{\phi_i(v_i)} %
\end{equation} 

\noindent where \textit{P$_c(v_i)$} is the probability of the 
\textit{i}-th star to be a cluster member. 
According to Bayes' minimum error rate decision rule,
 a threshold value of 0.5 minimises the misclassification.
 At the end of the analysis, 536 stars (out of 1028)
have a membership probability larger than 0.5. 
The mean RV of these stars is 35.9\,km\,s$^{-1}$, 
with a standard deviation of 2.8\,km\,s$^{-1}$.

Note that undetected member binaries may have discrepant 
RVs and be classified as non-members by this procedure.
Rigourously speaking, the RV distribution of the members
may deviate from a Gaussian. 
The measured RVs of unresolved binaries are the
combination of the motion of the center of mass and of the orbital
motion, and tend to have larger uncertainties than
isolated stars.
The result of convolving a binary model with a normal
distribution is a modification of the tails of the Gaussian, by
extending and increasing them. Here some unresolved pairs may therefore be classified as non-members.

\subsection{Membership of the UVES stars} \label{sec:membershipU}

\begin{figure}[ht!]
\begin{center} \resizebox{\hsize}{!}{\includegraphics[scale=0.57]{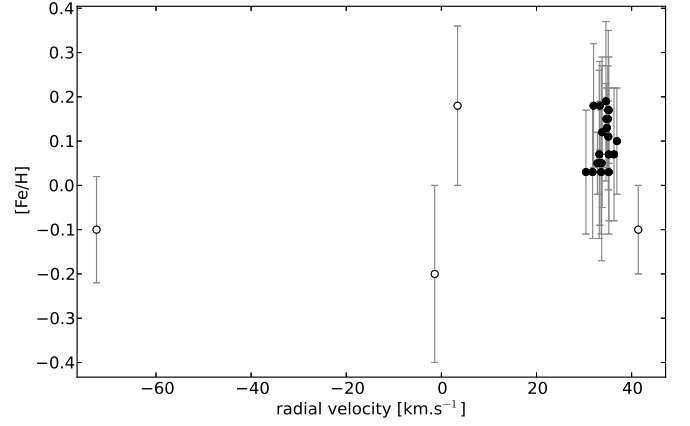}} \caption{\label{fig:radvelFeH} Identification of the red clump cluster members from radial velocities and iron abundance. The error bars for the radial velocities are smaller than the symbols. The 21 filled circles correspond to the members, while the 4 open circles are the stars we consider non-members.} \end{center}
\end{figure}

We have stellar parameters and accurate metallicities for the 25 UVES targets of NGC\,6705, 
including three stars that show a very discrepant RV and are not members of the
 clusters (see Fig.~\ref{fig:radvelFeH}). 
A fourth star (18510093-0614564) has an outlying radial velocity of 41.4\,km\,s$^{-1}$, even though its photometry is compatible with the other
 cluster members. As a possible explanation, this star could be a single-lined binary, made of a red clump star and a main-sequence star. Being much hotter, 
the main-sequence companion contributes to the spectrum with a continuum emission only,
 making the lines shallower. 
This hypothesis would explain why this star has a discrepant radial velocity, and is also an outlier in metallicity,
 with [Fe/H]$=-0.10 \pm 0.10$  (see following section).
 In the absence of further elements, we did not consider it as a member in the rest of this study.

Finally, 21 stars can be considered as bona-fide members of the cluster. 
The mean RV for the UVES members is 34.1\,km\,s$^{-1}$ (with a standard deviation of
 1.5\,km\,s$^{-1}$), which is lower than the mean value of 35.9$\pm$2.8\,km\,s$^{-1}$ found for GIRAFFE stars. The lack of targets in common between both instruments do not allow for a solid comparison of the systematics between UVES and GIRAFFE, and both results are compatible within their standard deviations.
In the first GES data release, \citet{Sacco14} note an average offset of 0.87\,km\,s$^{-1}$ between the UVES and GIRAFFE HR15N radial velocities, which is consistent with the offset we observe here.

\section{Extinction maps} \label{sec:extmap}

\begin{figure}[ht]
\begin{center} \resizebox{\hsize}{!}{\includegraphics[scale=0.6]{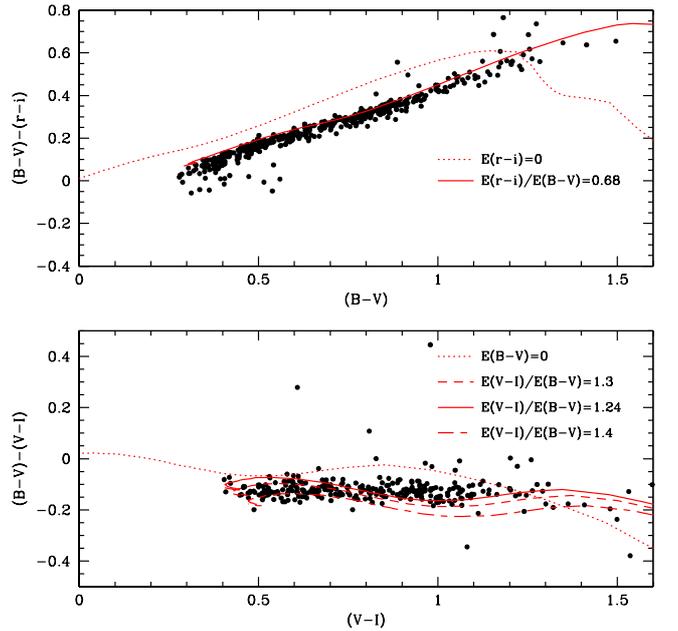}} \caption{\label{fig:extlaw} Color-color diagrams on NGC\,6705 GIRAFFE members main-sequence stars, compared with the relation expected from PARSEC isochrones for various extinction laws.} \end{center}
\end{figure}

When looking towards the inner parts of the Galaxy, the line of sight often meets regions of high extinction. 
Before discussing the age and the structure of NGC\,6705, we need to assess whether the studied region is affected by differential extinction.
The available VPHAS+ $ri$ photometry covers a very wide field, much larger than our $BVI$ photometry, which allows us to study the extinction of the background around the cluster.
The $ri$ CMD of the inner 12$\arcmin$ of the cluster is shown in Fig.~\ref{fig:VPHASCMD}. 
The PARSEC isochrone \citep{Bressan12} shown in that figure corresponds to the parameters of the best fit of the $BV$ photometry (see Sect.~\ref{sec:params}).
First we make use of the GIRAFFE members to derive the extinction law in the direction of the cluster.
Fig.~\ref{fig:extlaw} presents the color-color diagrams in the passbands $BVIri$. To fit the data, we need
to adopt the relations $E(V-I)=1.24 (\pm 0.05)\times E(B-V)$ and  $E(r-i)=0.68 (\pm 0.04) \times E(B-V)$.
 While the ratio $E(V-I)/ E(B-V)$ is consistent with the total-to-selective extinction ratio R$_V=A_V/E(B-V)$=3.1
using the \citet{Fitzpatrick99} law, the relation $E(r-i)/ E(B-V)$ is slightly larger than the value of
 $E(r-i)=0.6 \times E(B-V)$ given by \citet{Yuan13}.

Red clump stars in intermediate and old populations are often used in literature as distance and extinction indicators. The PARSEC isochrones indicate that the $r-i$ color of the red clump varies by 0.07 for ages from 2 to 10\,Gyr, and by 0.14 for Z from 0.001 to 0.03. Due to the small dependence of their magnitude and color on age and metallicity, they make good tracers of the extinction across the field. 

\begin{figure*}[ht]
\begin{center} \resizebox{\hsize}{!}{\includegraphics[scale=0.6]{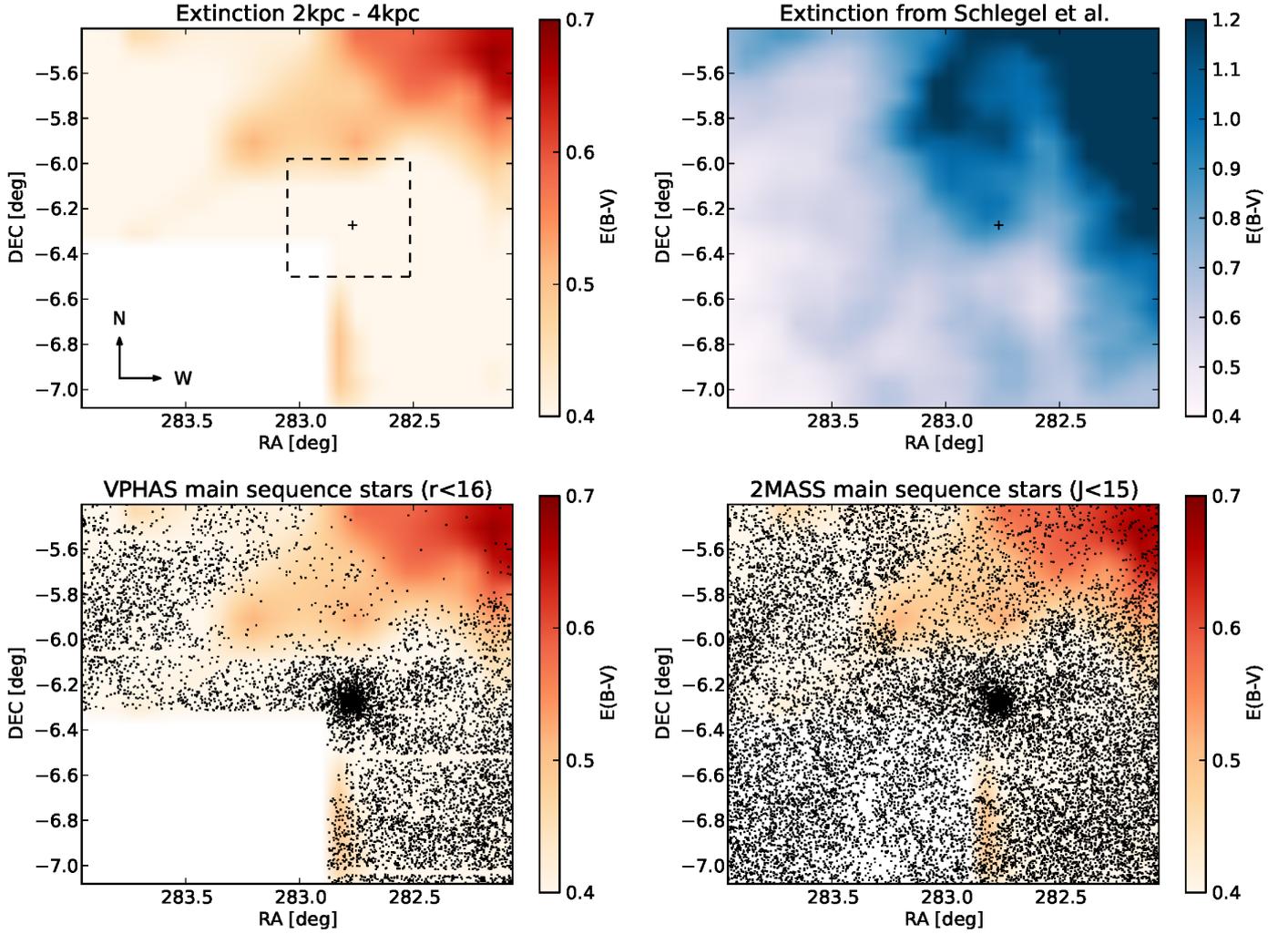}} \caption{\label{fig:extmapVPHAS}\textit{Top-left:} extinction map obtained from VPHAS+ photometry in the distance range 2-4\,kpc. The cross indicates the center of the cluster. The dashed line shows the footprint of our $BVI$ photometry. \textit{Bottom-left:} same extinction map. The black points are cluster stars selected from the $(r-i,r)$ VPHAS+ CMD, with $r$<16. The CCD gaps are visible as horizontal lines. \textit{Top-right:} extinction map from \citet{Schlegel98} for this field. \textit{Bottom-right:} same as bottom-left, but with 2MASS main-sequence stars ($J$<15).} \end{center}
\end{figure*}

We simulated a field corresponding to the coordinate of the cluster using the Padova Galaxy Model \citep{PadovaGalMod}. The red clump stars in the model have an absolute $r$ magnitude of $0.55$
and an intrinsic color $(r-i)=0.525$. 
We followed the color excess of the red clump stars at different distances. The field was divided in cells of $0.25\degr \times 0.25\degr$. In each of these cells, we computed the reddening as follows: the red giant stars falling in the magnitude range that corresponds to the distance range we want to probe (for instance, 11.5 < $r$ < 13 for 2 -- 4\,kpc) were divided in bins of 0.05 in the $(r-i)$ color, and we looked for the most populated color bin. This color was then compared to the reference value of $(r-i)=0.525$, to find the color excess. 
 We applied the procedure a second time, looking at a fainter magnitude slice, to take into account the fact that one must look at fainter magnitudes to probe stars that are more reddened. We found that going for more iterations did not significantly change the results. After this second step, the final color excess $E(r-i)$ was converted to $E(B-V)$ using the relation $E(r-i)=0.68 \times E(B-V)$. Our bins of 0.05 in $r-i$ color correspond to a resolution of 0.07 in $E(B-V)$. The extinction map we obtain for the distance range $2-4$\,kpc is visible in Fig.~\ref{fig:extmapVPHAS}.

The inner region of the cluster studied with the WFI $BVI$ photometry sits in a zone of relatively low and uniform extinction, with $E(B-V)=0.40$ out to $11\arcmin$ from the center. A higher extinction is found at DEC > -6\degr and RA < 283.25\degr, with $E(B-V)$ up to 0.7. A comparison with the map by \citet{Schlegel98} shows consistency, with the north-west quadrant being the most reddened. However, the maps of \citet{Schlegel98} give the integrated extinction along the line of sight while we restrict our determination
to the distance range 2--4\,kpc. For this reason, our values of extinction are lower and the distribution is not directly comparable.

Although the cluster itself lies in a window of constant extinction, the bottom-left panel of Fig.~\ref{fig:extmapVPHAS} shows clearly that the star density distribution of the photometrically selected VPHAS+ cluster stars (Fig.~\ref{fig:VPHASCMD}) is correlated with the extinction, making the background density inhomogeneous.

\begin{figure}[ht]
\begin{center} \resizebox{\hsize}{!}{\includegraphics[scale=0.6]{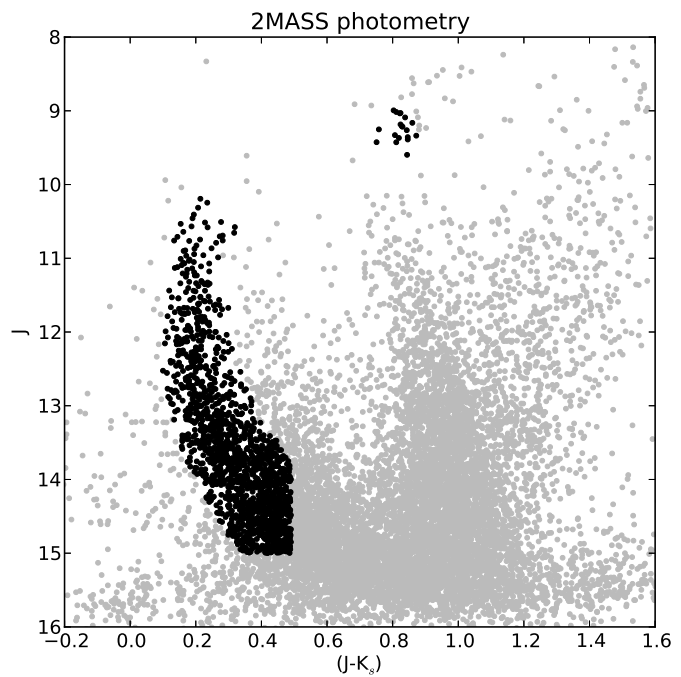}} \caption{\label{fig:2MASSCMD}2MASS CMD of the inner 12$\arcmin$ of NGC\,6705. The spatial distribution corresponding to the photometric selection (black points) is shown in Fig.~\ref{fig:extmapVPHAS} on a larger field.} \end{center}
\end{figure}

In the bottom-right panel of Fig.~\ref{fig:extmapVPHAS} we show the distribution of photometrically selected stars from the 2MASS catalog with $J$ < 15 (Fig.~\ref{fig:2MASSCMD}). The infrared bands of the 2MASS data are less affected by extinction than the $r$ and $i$ bands of VPHAS+, and the relations by \citet{Yuan13} tell us that a reddening of $E(B-V)=0.7$ translates to an extinction of $0.54$\,mag in the $J$ band. The distribution obtained for the 2MASS stars is similar to the one obtained for the VPHAS+ stars, with a hole in the distribution in the north-east quadrant and a steep drop in the stellar density in the north-west.

Further details on this density distribution are given when we establish the luminosity profile of the cluster (Sect.~\ref{densityprof}).

\section{Cluster structure} \label{sec:structure}
To derive the cluster structure, we make use of the WFI photometry in the inner region, and of the 
VPHAS+ data in the external parts.

\subsection{Center determination} \label{sec:centerDet}
To determine the position of the center of the cluster, we first made a photometric selection that includes the brightest stars (red clump and upper main sequence). We used the method described by \cite{Donati12} that consists in finding the barycenter of the sample, then retaining the 70\% of the stars that lie closest to this barycenter and repeat the procedure until convergence to the actual center of the cluster. The uncertainty on the final position arguably depends on the uncertainty in the selection of cluster members. Selecting only the brightest stars gives a better-defined membership but also poorer statistics. Using different magnitude cuts from V=13 to 16, the position of the center is at: $\alpha_{2000}=18^{h} 51^{m} 4^{s}$, $\delta_{2000}= -6\degr 16\arcmin 22\arcsec$ (RA=$282.767\degr$,DEC=$-6.273\degr$) with an uncertainty of 2\arcsec.

\subsection{Radial density profile}\label{densityprof}
 We derived the luminosity profile using the $V$-band magnitudes. Through a $\chi^{2}$ minimisation, we fitted a single mass 2-parameter King profile \citep{King62}:

\begin{equation} \label{kingProf}
f(r) = \sigma_{bg} + \frac{ \sigma_{0} }{ 1+(r/r_{core})^{2} }
\end{equation}

\noindent where $\sigma_{bg}$ the background luminosity, $\sigma_{0}$ the central luminosity and $r_{core}$ the core radius are left as free parameters.

\begin{figure}[ht]
\begin{center} \resizebox{\hsize}{!}{\includegraphics[scale=0.55]{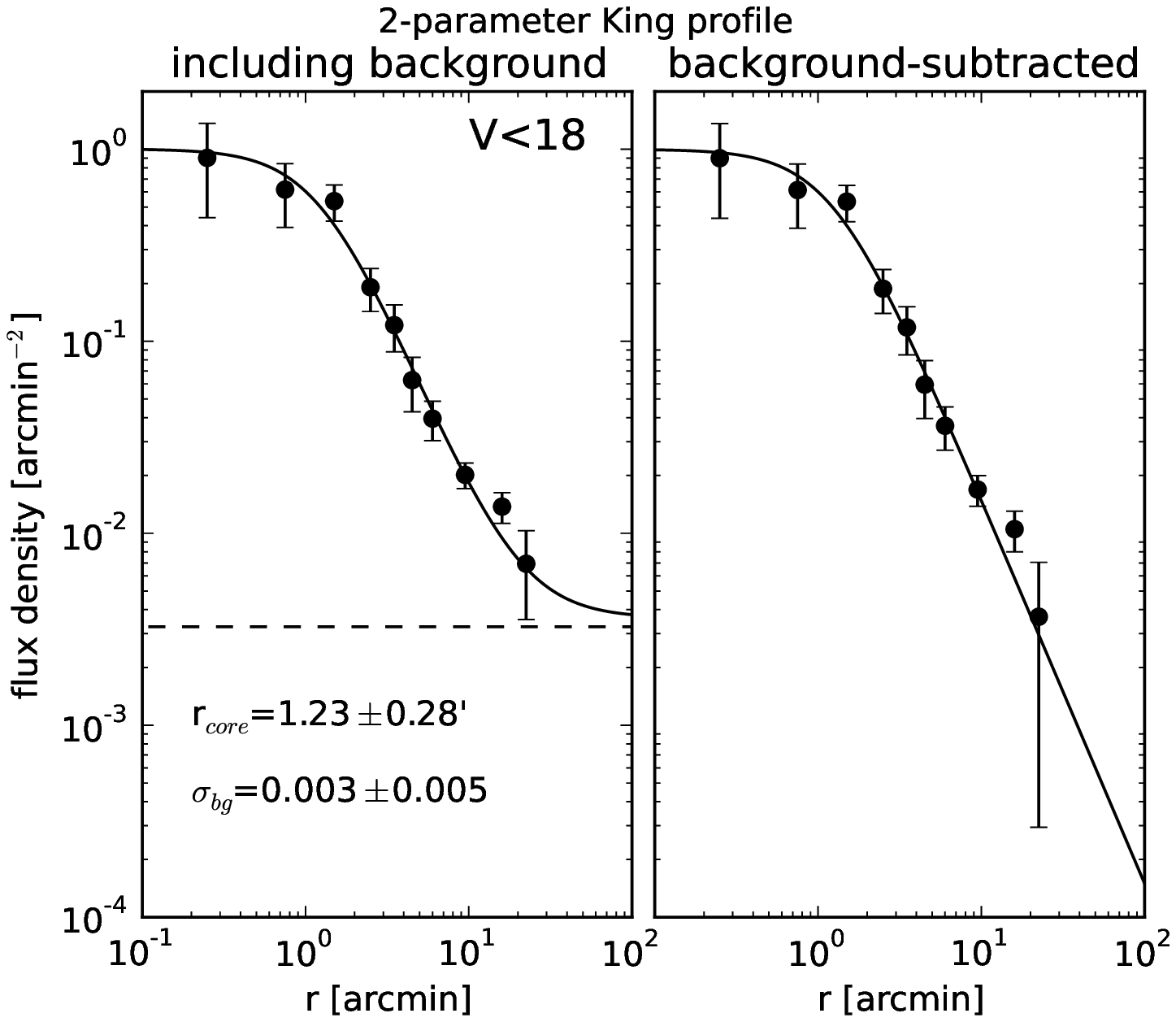}} \caption{\label{lumprof} Observed luminosity profile (black dots) for stars with $V$<18, fitted with a two-parameters King profile (continuous line), including background (left) and after subtracting the background (right). The flux density was normalised to the central luminosity so that $\sigma_{0}=1$.} \end{center}
\end{figure}

The data was corrected for completeness in each radius and magnitude bin.
Using all the stars with $V$<18, we obtain a core radius of $1.23 \pm 0.28\arcmin$ (Fig.~\ref{lumprof}). This value is in agreement with the value of $1.81 \pm 0.71\arcmin$ found by \citet{Santos05} (hereafter S05) using main-sequence stars brighter than $J$=15. Taking into account our distance estimate (Sect.~\ref{params}) we obtained a core radius of $0.69 \pm 0.24$\,pc in physical units. 

\begin{figure}[ht]
\begin{center} \resizebox{\hsize}{!}{\includegraphics[scale=0.55]{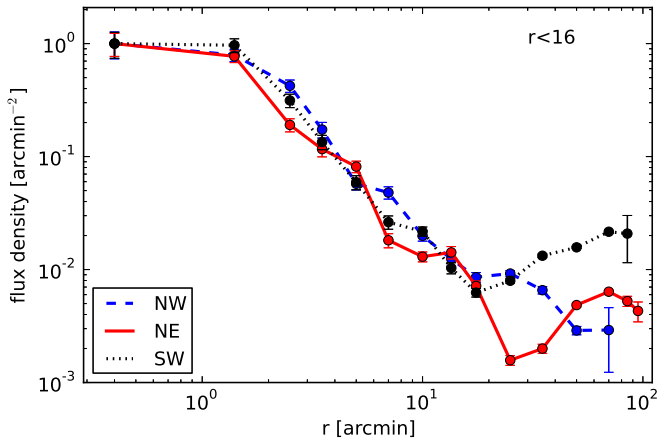}} \caption{\label{extlumprof}Stellar density profiles for the north-west, north-east and south-west quadrants (respectively NW, NE and SW) of the VPHAS+ photometry, using stars with $r<16$. The profiles in the three quadrants were normalised to the flux in the innermost bin. The background level was not subtracted.} \end{center}
\end{figure}

The much wider field offered by the VPHAS+ data enables us to trace the luminosity profile in the most distant parts of the cluster.
We established the luminosity profiles in three different quadrants (Fig.~\ref{extlumprof}). We can see that the three profiles are similar inside 10$\arcmin$, with the small differences being explained by statistics and a patchy extinction, but show different behaviours at larger radii. As one could expect from the extinction map of Fig.~\ref{fig:extmapVPHAS}, the luminosity in the north-west quadrant drops to lower values than in the other two. In the north-east quadrant, a dip is visible between 25$\arcmin$ and 40$\arcmin$, corresponding to a region of stronger extinction of the background (see Fig.~\ref{fig:extmapVPHAS}). In the south-west quadrant, beyond 20$\arcmin$, the background density increases with radius as the extinction decreases. Since outside of the core the luminosity profile is mainly shaped by extinction, it is not possible to estimate the tidal radius of NGC\,6705 by fitting a 3-parameter King profile. The value of $r_{tidal}=52\pm27\arcmin$ reported by S05 falls exactly in the distance range affected by strong extinction. S05 also observed a deviation of the luminosity profile from their model between 11$\arcmin$ and 16$\arcmin$, and suggest this excess could be due to low-mass stars moving to larger radii due to mass segregation. Without a proper model for the background density, it seems very difficult to draw conclusions on the structure of the cluster beyond 10$\arcmin$.

\subsection{Mass segregation from the luminosity profile} \label{sec:massseg}
Mass segregation is an internal dynamical phenomenon that takes place in star clusters and leads the most massive stars to be more tightly clumped than the least massive ones \citep{Spitzer69}. A certain number of methods can be used to see whether or not mass segregation is occurring in a cluster, for instance applying a Minimum Spanning Tree analysis \citep{Allison09}, or directly comparing the distance to the center for stars of different brightness. Here we compare the density profiles of stars in different mass ranges (cf. Sect.~\ref{densityprof}) and compare the slope of the mass function in different regions of the cluster (our Sect.~\ref{massfunct}, and S05). 

\begin{figure}[ht]
\begin{center} \resizebox{\hsize}{!}{\includegraphics[scale=0.48]{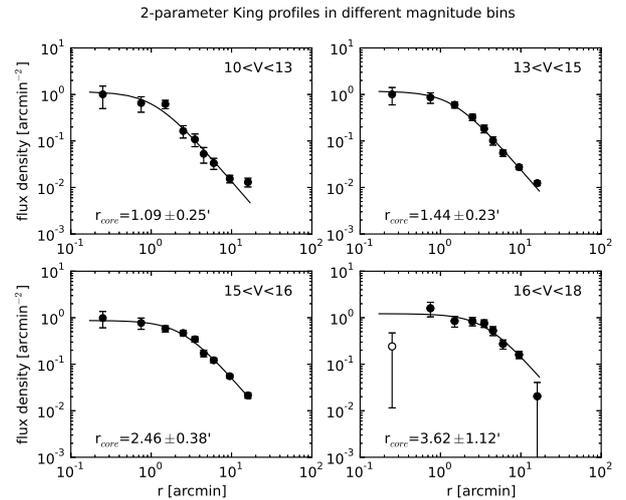}} \caption{\label{4kings}Observed and fitted luminosity profiles obtained with stars in different magnitude ranges. The background was fitted and subtracted. In the bottom-right panel, the open symbol is the datapoint that was discarded when fitting the luminosity profile.} \end{center}
\end{figure}

Figure~\ref{4kings} shows the luminosity profiles established in four different magnitude ranges: 10 -- 13, 13 -- 15, 15 -- 16, 16 -- 18. For our best-fitting PARSEC isochrone, these correspond to the mass ranges: M > 2.14\,M$_{\sun}$, 2.14 -- 1.37, 1.37 -- 1.14, 1.14 -- 0.81. For stars with $V>16$, we applied a completeness correction based on the distance and magnitude.

In the faintest range ($16<V<18$) we discarded the data point corresponding to the most central bin ($r<0.5\arcmin$), due to the very low completeness for faint stars at the center of the cluster. It is clear that the core radius increases as fainter stars are considered, from $1.02 \pm 0.25\arcmin$ in the range $10 < V < 13$ to $3.62 \pm 1.12\arcmin$ in the range $16 < V < 18$. This result is consistent with results by \citet{Sung99}, although they do not quantify the increase in radius for the distribution of progressively fainter stars.

\section{Kinematics} \label{sec:kinematics}
\subsection{Rotation}
\begin{figure}[ht]
\begin{center} \resizebox{\hsize}{!}{\includegraphics[scale=0.6]{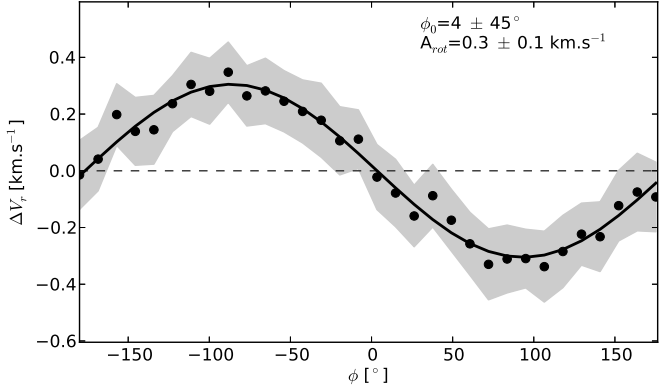}} \caption{\label{fig:rotation}Rotation curve obtained from the cluster members. The shaded region show the uncertainty estimated via bootstrapping. The best fitting sine corresponds to a position angle of the rotation axis $\phi_{0}=4\degr$ and an amplitude $A_{rot}=0.3$\,km\,s$^{-1}$.} \end{center}
\end{figure}

Since NGC\,6705 is a massive system and radial velocities are available for a large number of targets, we looked for hints of the rotational signature of the cluster, as is routinely done for GCs \citep[see e.g.][]{Cote95,Bellazzini12,Bianchini13}.
We first corrected the GIRAFFE radial velocities for the offset observed in Sect.\ref{sec:dataHARPS}, then applied the procedure to the whole sample. The method consists in dividing the cluster in two and selecting one half, according to a position angle $\phi$ measured here from north to east ($\phi=0\degr$ corresponds to the north half, while $\phi=+90\degr$ and $\phi=-90\degr$ correspond to the east and west halves, respectively). The mean radial velocity is computed in the selected region. Varying the angle $\phi$ by steps of 5$\degr$ we trace the mean radial velocity in different regions.
For a rotating object, we expect the radial velocities to follow a sinusoidal dependance on the position angle $\phi$, being equal to the mean radial velocity of the cluster when $\phi$ corresponds to the orientation of the rotation axis (projected on the sky), following the equation:

\begin{equation} 
\Delta V_r= A_{rot}sin({\phi}_{0}+\phi)
\end{equation}

\noindent where $A_{rot}$ corresponds to the amplitude of the rotation and $\phi_{0}$ is the position angle of the rotation axis.
We estimated the uncertainty on the mean radial velocity by bootstrapping: in each position angle bin, containing N datapoints, we created 100 sets of N datapoints by randomly sampling the radial velocity dataset, calculated the mean velocity for each one, and computed the standard deviation of these results. The best-fitting sine function corresponds to an orientation angle $\phi_{0}=4\pm45\degr$, and an amplitude $A_{rot}=0.3\pm0.1$\,km\,s$^{-1}$ around the mean radial velocity (Fig.~\ref{fig:rotation}).
Since the RV is averaged
over the full range of radii covered by the sample, the derived  A$_{rot}$
is a lower limit to the maximum rotation amplitude, depending on the rotational gradient of the cluster. The observed pattern is however very weak, with an amplitude much smaller than the velocity dispersion of the cluster, and we can conclude that rotation does not play a significant role in the observed velocity dispersion in NGC\,6705.

\subsection{Intrinsic velocity dispersion} \label{sec:intrinsic}
Due to uncertainties on the radial velocities of each invididual star, the velocity dispersions we observe in our samples (2.8\,km\,s$^{-1}$ for the GIRAFFE sample, 1.5\,km\,s$^{-1}$ for the combined HARPS and UVES sample) are larger than the true, intrinsic dispersion of those stars. Assuming this intrinsic distribution is gaussian, and that the quoted uncertainties $\left\{\epsilon_1,...,\epsilon_N\right\}$ on the observed radial velocities $\left\{v_1,...,v_N\right\}$ are gaussian errors, it is possible to estimate the dispersion $\sigma$ and the mean radial velocity $\mu$ of the true underlying distribution \citep[see e.g.][]{Walker06}.

The probability of measuring the radial velocity $v_i$ by sampling a random point from a normal distribution that has mean $\mu$ and dispersion $\sigma$ is:

\begin{equation}
 P(v_i,\mu,\sigma) = \frac{1}{\sqrt{2 \pi \left( \sigma^2 + {\epsilon_{i}}^2\right)} } \exp \left( - \frac{1}{2} \frac{ (v_i - \mu)^2 }{\sigma^2 + {\epsilon_{i}}^2} \right)
\end{equation} 

For a sample of N stars, the joint probability is simply the product of the N individual probabilities:

\begin{equation}
 \mathcal{L} = P(\left\{v_1,...,v_N\right\},\mu,\sigma) = \prod_{i=1}^N \frac{1}{\sqrt{2 \pi \left( \sigma^2 + {\epsilon_{i}}^2\right)} } \exp \left( - \frac{1}{2} \frac{ (v_i - \mu)^2 }{\sigma^2 + {\epsilon_{i}}^2} \right)
\end{equation} 

\noindent and the best estimates of $\mu$ and $\sigma$ are those that maximise $\mathcal{L}$. Applying this method to our sample of GIRAFFE RVs, we found the values $\mu_{G}=36.0\pm0.2$\,km\,s$^{-1}$, $\sigma_{G}=2.5\pm0.1$\,km\,s$^{-1}$. For the HARPS+UVES sample the results are: $\mu_{H+U}=34.8\pm0.4$\,km\,s$^{-1}$, $\sigma_{H+U}=1.3^{+0.3}_{-0.2}$\,km\,s$^{-1}$. 
We do not expect such large difference between the radial velocity dispersion of both samples. The most likely source of this difference is that the quoted uncertainties on the GIRAFFE radial velocities are underestimated. The GIRAFFE targets are hot, fast-rotating stars for which RVs can be difficult to measure, and the main source of dispersion in the GIRAFFE sample is likely to be the measurement errors. This hypothesis may be tested when the final analysis is available for all gratings in a further GES data release.

\subsection{Virial mass}\label{virmass}
A complete kinematical study of the cluster would require proper motions and radial velocities for a significant share of the stars. We do not have such a dataset, but we have accurate radial velocities for 31 red clump stars of NGC\,6705 (combining the HARPS and UVES datasets) that we can use to estimate a lower limit to the total mass of the cluster.

NGC\,6705 exhibits mass segregation, which is a sign of a dynamically relaxed cluster, at least in its core. The mass of a dynamically relaxed cluster is related to its average radius and velocity dispersion through the virial theorem:

\begin{equation}
   M_{tot} = \frac{2 \langle v^2 \rangle \langle R \rangle}{G} %
\end{equation} 

\noindent where $G$ is the gravitational constant, $\langle R \rangle$ the average radius, and  $\langle v^2 \rangle = v_{\mu}^{2} + v_{\sigma}^{2} + v_{r}^{2}$ the 3-dimensional velocity dispersion. In this study we only have access to the radial velocity dispersion $v_{r}$. However, assuming that the velocity distribution is isotropic, the three components of the velocity dispersion are equal and we can write:

\begin{equation}
   \langle v^2 \rangle  =  v_{\mu}^{2} + v_{\sigma}^{2} + v_{r}^{2} =  3 v_{r}^{2} 
\end{equation} 

With $\sqrt{\langle v_{r}^{2}\rangle} = 1.34^{+0.32}_{-0.22}$\,km\,s$^{-1}$ and an average radius $\langle R \rangle = 3.42\arcmin$ (2.0$\pm$0.2\,pc) (with the most distan star locate at an agular distance of $10.1\arcmin$) this equation yields: M$_{tot}=5083\pm1600$\,M$_{\odot}$. The number is in good agreement with the virial mass obtained by \citet{McNamara77} with the same method, but using proper motions ($5621$\,M$_{\odot}$). It is possible to plug in the values obtained from the GIRAFFE targets, that spread out to $12.6\arcmin$ and show a larger average radial distance and a considerably higher velocity dispersion (4.52$\arcmin$ and 2.8\,km\,s$^{-1}$, respectively). The result is a mass of $21\,900\pm2700$\,M$_{\odot}$. Since the velocity dispersion of the GIRAFFE stars is affected by the larger uncertainty on the radial velocities, this value can be considered an upper limit.

This simple method only provides a first order estimation. A  more sophisticated  approach would include
detailed kinematical modelling,  for instance adopting a multi-mass model approach as is generally done for GCs \citep[see for instance][]{Miocchi06}. This is however, outside the scope of the paper.
The calculation presented here only considers the positions and radial velocities of the massive stars in the central region, and may suffer from biases due to the low number of selected targets. With a mass of the order of $5\times10^{3}$\,M$_{\odot}$ contained in the inner part (within 2\,pc), we confirm that NGC\,6705 is a relatively massive OC.

\section{Age determination} \label{sec:params}
In this section we give an estimate of the age, distance, and metallicity of NGC\,6705 from the CMD analysis of the WFI photometry, which corresponds to the part of the cluster that is not affected by variable extinction.
We first compared the CMD with PARSEC \citep{Bressan12} and Dartmouth \citep{Dotter08} isochrones, then proceeded to fit a synthetic luminosity function in the inner region of the cluster.

\subsection{Comparison with theoretical isochrones} \label{params}
We have compared the $BVI$ CMD of the inner 6$\arcmin$ of NGC\,6705 with PARSEC and Dartmouth isochrones. The age, distance modulus, reddening and metallicity of the cluster can be estimated by matching the key features of the CMD (such as the main sequence slope and the positions of the main sequence turn-off and the red clump) traced by the cluster members established in Sect.~\ref{sec:membershipG} to the shape of the theoretical isochrones.

\begin{figure*}[ht]
\begin{center} \resizebox{\hsize}{!}{\includegraphics[scale=0.5]{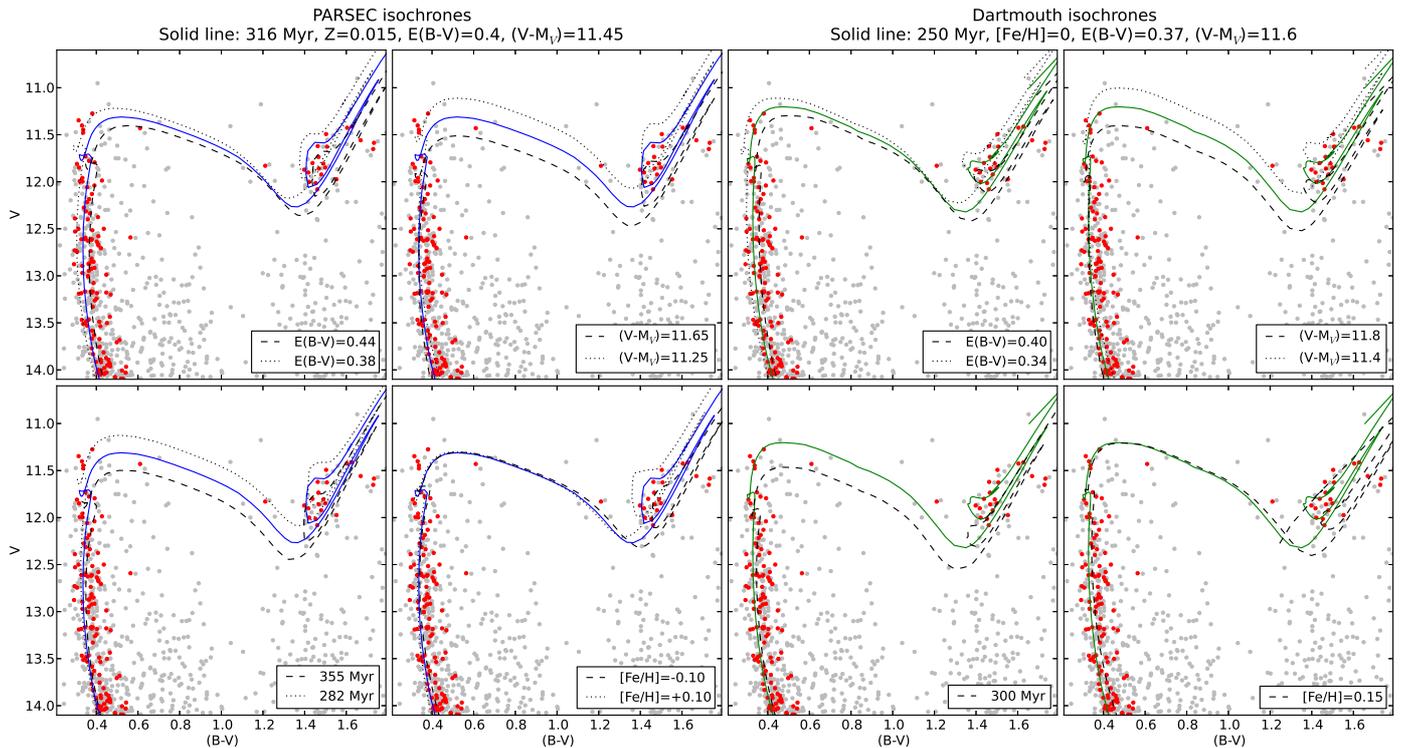}} \caption{\label{fig:8isos}The grey points are all the stars in our photometry, and the radial velocity members are marked in red. \textit{Left-hand side panels:} The best-fitting PARSEC isochrone (solid blue line) corresponds to an age of 316 Myr, a solar metallicity, a reddening $E(B-V)=0.40$ and a distance modulus $(V-M_{V})=11.45$. The four panels show the change in the isochrone when modifying the extinction, distance modulus, age and metallicity with respect to the best-fitting isochrone. \textit{Right-hand side panels:} same procedure with Dartmouth isochrones.  The best-fitting isochrone is the solid green line. In all cases, the data are uncorrected for extinction (the isochrones are shifted along the extinction vector). } \end{center}
\end{figure*}

The PARSEC isochrone that reproduces best the morphology of the CMD is a model of 316\,Myr (log $t$=8.5), solar metallicity (Z=0.0152), $E(B-V)=0.40$ and $(V-M_{V})=11.45$ (d$_{\sun}=1995 \pm 180\,pc$). The left-hand side panels of Fig.~\ref{fig:8isos} shows this isochrone, along with different choices of distance modulus and extinction, and different ages and metallicities. 
Using a PARSEC isochrone of slightly super-solar metallicity produces a marginally compatible fit. On the other hand, a sub-solar metallicity fails to reproduce the position of both the red clump and the main sequence turn off at the same time. 
We estimate the uncertainty on the extinction to be of the order of $0.03$, the uncertainty on the age $50\,$Myr. 
For an age of 316\,Myr, the mass of the main sequence turn off stars is $3.2\,\text{M}_{\odot}$.

The best set of parameters for fitting a Dartmouth isochrone is an age of 250\,Myr, extinction $E(B-V)=0.37$, distance modulus $(V-M_{V})=11.6$ (d$_{\sun}$=2090\,pc), shown in Fig.~\ref{fig:8isos}. Again, using isochrones of super-solar metallicities produces a less satisfying fit than a solar metallicity. The age of 250\,Myr is in agreement with the result of S05, who used Dartmouth isochrones on 2MASS photometry. This age corresponds to a turn off mass of $3.47\,\text{M}_{\odot}$.

\begin{figure}[h]
\begin{center} \resizebox{\hsize}{!}{\includegraphics[scale=1.0]{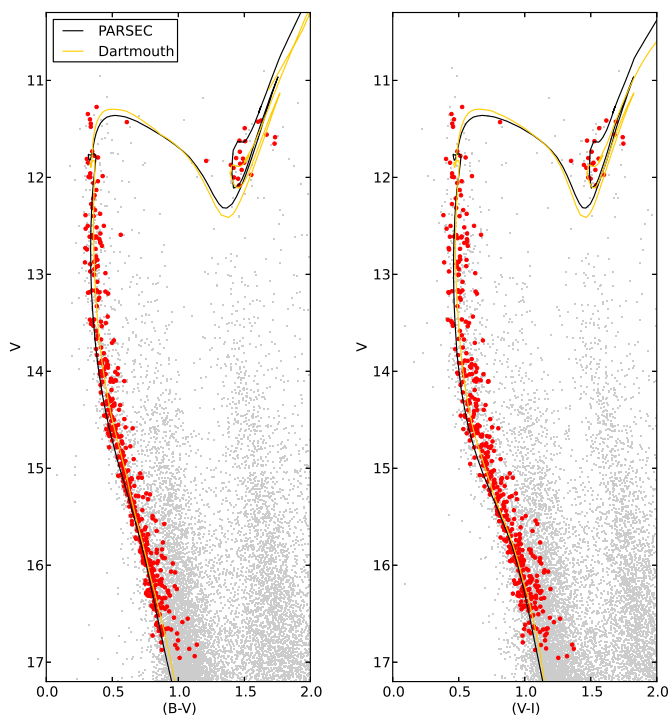}} \caption{\label{padovaVSdartmouth} Comparison between theoretical isochrones and our photometry (PARSEC isochrone of solar metallicity, 316\,Myr, shifted by $E(B-V)=0.40$ and $(V-M_{V})=11.45$) and Dartmouth isochrone of solar metallicity, 250\,Myr, $E(B-V)=0.37$, $(V-M_{V})=11.6$). The grey points are all the stars in our photometry, while the red points are the radial velocity members.} \end{center}
\end{figure}

The best fits for these two sets of isochrones are both shown on Fig.~\ref{padovaVSdartmouth} for a direct comparison.

As already discussed, PARSEC and Dartmouth isochrones give different determination of the age.
This difference is due to the fact that the red clump stars tend to be brighter in the PARSEC tracks than in the Dartmouth tracks, because of a different choice of mixing-length parameter and solar metallicity reference \citep[see Sect.~5.6 of][]{Bressan12}. As a consequence, a PARSEC isochrone of higher age is necessary to reproduce the position of the red clump.

In order to fit the $(V-I)$ color in Fig.~\ref{padovaVSdartmouth} we adopt the 
relation $E(V-I)/E(B-V)=1.24$ derived in Section \ref{sec:extmap}.

\begin{figure}[ht]
\begin{center} \resizebox{\hsize}{!}{\includegraphics[scale=0.50]{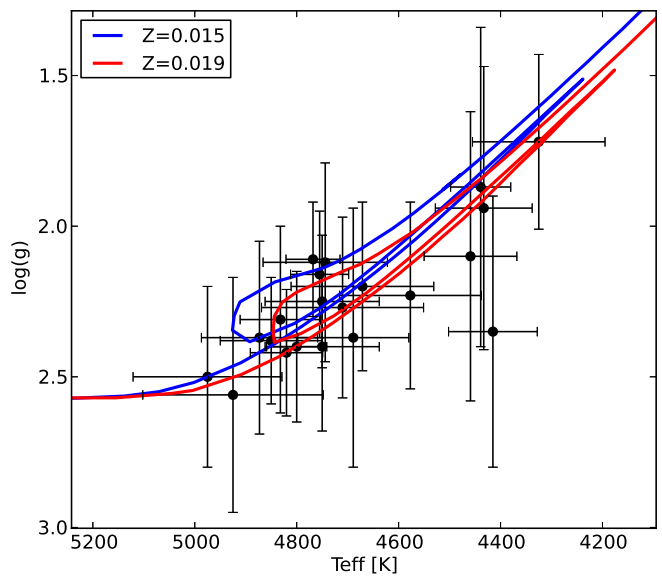}} \caption{\label{fig:Tefflogg}
 Position of the 21 UVES members in the theoretical plane. 
The position of the red clump is well reproduced with a
 PARSEC isochrone of age log $t$=8.5 and Z=0.015 ([Fe/H]=0) or Z=0.019 ([Fe/H]=0.1).} \end{center}
\end{figure}

 Fig.~\ref{fig:Tefflogg} shows the position of 
the 21 UVES cluster members in the (T$_{\textrm{eff}}$ -- log~$g$) plane.
 PARSEC isochrones of age 316\,Myr and solar or slightly super-solar metallicity 
are compatible with the location of the red clump stars.

\subsection{Luminosity function}

To confirm the results obtained from the comparison with theoretical isochrones, we have studied the luminosity function (LF) of the inner $0.1^{\circ}$ ($6\arcmin$) of the cluster. This region is large enough to provide good statistics, without being too contaminated by background stars. A photometric selection following the main sequence down to magnitude $V$=18 was made. We considered the stars further than $0.3^{\circ}$ ($18\arcmin$) from the center of the cluster as our background field (see Fig.~\ref{cmd1}). The LF of the background field was subtracted from the LF of the cluster, taking into account the respective completeness and area of the two regions. The error on the luminosity function takes into account the statistical error on both the cluster and the background. 

Synthetic populations were computed using PARSEC tracks \citep{Bressan12}, varying the age and metallicity, as well as the slope $\alpha$ of the initial mass function (IMF) and the distance modulus. The reddening was kept fixed to a value of $E(B-V)=0.40$. The LF of the synthetic populations were compared with the observed LF using a $\chi^{2}$ method, imposing a constrain on the color of the main sequence and red clump.

\begin{figure*}[ht]
\begin{center}
\includegraphics[scale=0.49]{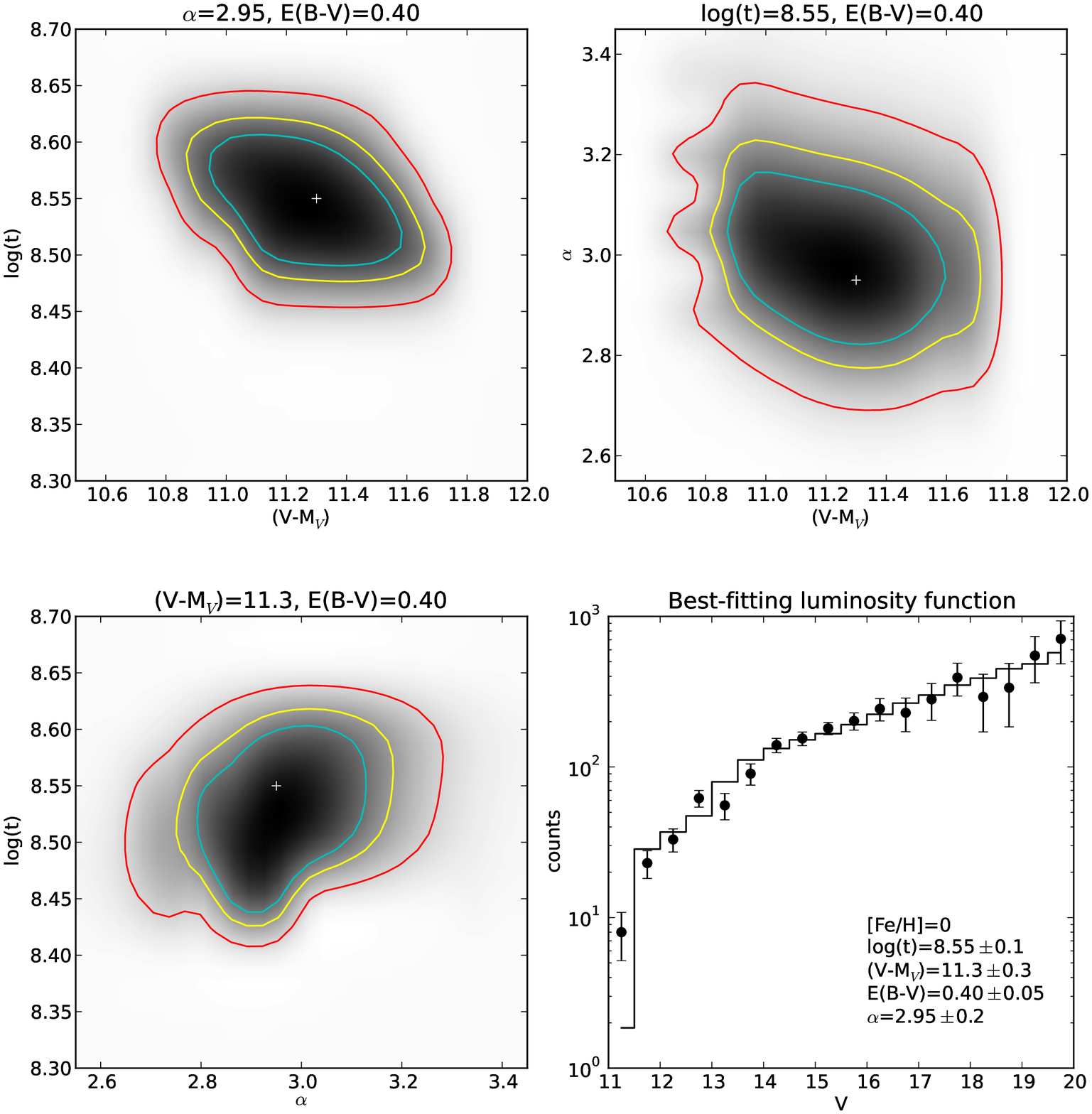} \caption{\label{fig:chi2maps}Probability contours for the fit of the luminosity function. The cyan, yellow, and red lines show the 50\%, 68\% and 90\% confidence regions, respectively. The white cross indicates the best-fitting solution. The bottom-right panel shows the observed LF (cleaned from the background) and the best-fitting theoretical one. The bins with $V>18$ are shown but were not used in the fitting procedure.}
\end{center}
\end{figure*}

Using isochrones of solar metallicity provide the lowest $\chi^{2}$, and isochrones with [Fe/H]= $+0.10$ provide very similar results, while using [Fe/H]= $-0.10$ provides less good fits. The $\chi^{2}$ maps (Fig.~\ref{fig:chi2maps}) show the confidence interval of the parameters that reproduce  the observed LF. From these maps we estimate the parameters of the cluster as: log $t$=$8.55 \pm 0.1$, $(V-M_{V})=11.55 \pm 0.3$, and the IMF slope $\alpha=2.95 \pm 0.2$.

In this procedure, we did not take into account that some of the stars may be unresolved binary (or multiple) systems. \citet{Weidner09} indicate that even in the extreme case where 100\% of the stars are in multiple systems, the $\alpha$ parameter may be underestimated by 0.1 at most.

\subsection{Mass function and total mass} \label{massfunct}

In this section we derive the mass of the cluster using its LF.
We produce the LF of the cluster in three annuli (corresponding to distance the ranges $0\arcmin-3\arcmin$, $3\arcmin-6\arcmin$, and $6\arcmin-9\arcmin$) and convert the $V$ magnitudes to masses using a PARSEC isochrone with the parameters determined in section \ref{params}. The observed magnitude range ($V$<18) corresponds to a mass range of 1 to 3.2 M$_{\sun}$. We have chosen to limit ourselves to the inner $9\arcmin$. At larger radii, the cluster is hardly visible against the field stars and our procedure that consists in subtracting the LF of the background from the observed one becomes very uncertain. 

\begin{figure}[ht]
\begin{center} \resizebox{\hsize}{!}{\includegraphics[scale=0.48]{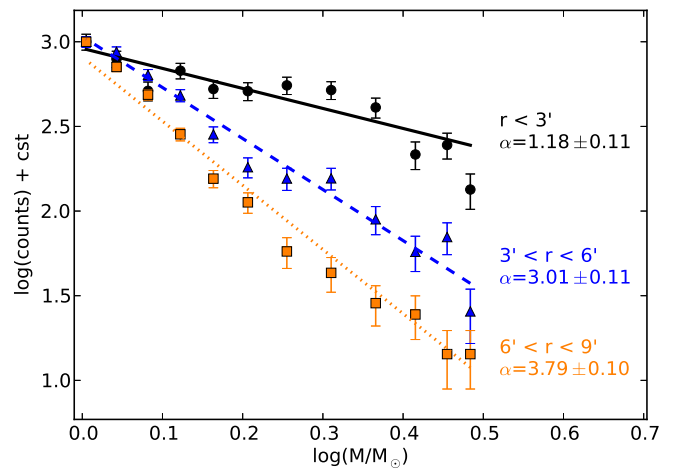}} \caption{\label{3MF}Mass functions obtained in three different regions of the cluster, showing the flattening of the IMF in the center. The y-axis was shifted so that the observed number of stars in the lowest mass bin is the same for all three.} \end{center}
\end{figure}

In each region, we have derived the parameter $\alpha$, defined as $N(M)\,dM\propto M^{-\alpha}$.
The results of the fit are shown in Fig.~\ref{3MF}. The value of $\alpha$ increases with the radius, indicating that the mass function gets steeper because of a deficit of high-mass stars. In the most central region, $\alpha=1.18$ indicates an excess of high-mass stars with respect to what would be expected from the Salpeter IMF ($\alpha=2.35$). Over the whole range 0 -- 9$\arcmin$, the slope of the mass function is $\alpha=2.70\pm0.2$.

The previous studies of the evolution of the slope of the IMF with the radius also found that the mass function was flatter in the inner region. S05 found a slope of $\alpha=0.27$ in the inner $1.8\arcmin$, $\alpha=2.41$ in the region 1.8 -- 10$\arcmin$, and $\alpha=3.88$ in the region 10 -- 21$\arcmin$. \citet{Sung99} found slopes of $\alpha=0.5$ to $1.7$ inside $3\arcmin$. The slope values found by those two studies are listed in Table~\ref{tab:MFslopes} along with our estimates. The distance ranges given by S05 were converted back from parsecs to arcminutes using their value of the distance.

\begin{table}
\begin{center}
	\caption{ \label{tab:MFslopes} Slope of the mass function in different regions on the cluster.}
	\small\addtolength{\tabcolsep}{-1pt}
	\begin{tabular}{l l}
	  \hline
	  \hline
	  Radius (\arcmin) & \multicolumn{1}{c}{$\alpha$} \\
	  \hline
	  \multicolumn{2}{c}{This study} \\
	  \hline
	  0 - 1.8 & 0.50 $\pm$ 0.15 \\
	  0 - 3   & 1.18 $\pm$ 0.11 \\
	  3 - 6   & 3.01 $\pm$ 0.11 \\
	  6 - 9   & 3.79 $\pm$ 0.10 \\
	  0 - 9   & 2.70 $\pm$ 0.19 \\
	  1.8 - 9 & 3.29 $\pm$ 0.07 \\
	  \hline
	  \multicolumn{2}{c}{Santos et al. (2005)} \\
	  \hline
	  0 - 1.8 & 0.27 $\pm$ 0.15 \\
	  1.8 - 10 & 2.41 $\pm$ 0.10 \\
	  10 - 21 & 3.88 $\pm$ 0.20 \\
	  0 - 21 & 2.49 $\pm$ 0.09 \\
	  \hline
	\end{tabular}
\tablefoot{$\alpha$ is the parameter in the mass function: $N(M)\,dM\propto M^{-\alpha}$.}
\end{center}
\end{table}

In theory, if the mass function of a cluster is known for its brightest stars, it is possible to extrapolate the mass functions down to smaller masses and estimate the total stellar mass contained in the cluster.
 When dealing with mass segregated  OCs possibly affected by evaporation and tidal mass loss, inferring the mass function from the observed range to the lower masses is affected by large uncertainties.
To obtain the slope of the mass function in the non-observed range (under 1\,M$_{\sun}$) we have used two methods. The simplest one is to assume that the slope of the power-law derived from the stars with M<1\,M$_{\odot}$ is the same all over the mass range. The second one is to use the IMF by \citet{Kroupa01}: $\alpha=1.3$ for $0.08<\text{M}_{\odot}<0.5$ and $\alpha=2.3$ for $0.5<\text{M}_{\odot}<1$. For masses over 1\,M$_{\sun}$ we have used the values obtained from our fit. In the inner region (r<$3\arcmin$) we observe a nearly flat mass function, which means that using this slope over the whole mass range produces fewer low-mass stars than using the Kroupa IMF. In the other two regions, the mass function is steeper and in the low-mass range it is well over the prediction of the Kroupa IMF. The result of integrating these mass functions are shown in Table~\ref{tab:mass} for the different choices of mass function in the different regions.

When integrating the mass function over the whole $0 - 9\arcmin$ region, we obtain values between 3683 $\pm$ 1063\,M$_{\sun}$ (using the Kroupa IMF under 1\,M$_{\sun}$) and 6851 $\pm$ 1865\,M$_{\sun}$ (using the extrapolated power-law). This latter number is compatible with the number quoted by S05, who estimate a total mass of 6500 $\pm$ 2100\,M$_{\sun}$ inside $10\arcmin$, and 11\,000 $\pm$ 3800 M$_{\sun}$ inside $21\arcmin$ using the IMF slopes listed in Table~\ref{tab:mass}, and the Kroupa IMF under 1\,M$_{\sun}$). 
Our determination does not take into account the presence of binary stars whose percentage is unknown. The effect of the presence of unresolved binaries on the observed $\alpha$ parameter is small, but according to \citet{Weidner09} not accounting for the presence of multiple systems can hide 15 to 60\% of the stellar mass of a cluster. In addition, less massive stars could be either lost from the cluster due to the effect of the disk tidal field, or located in a surrounding halo. For these reasons, our determination is a lower limit to the cluster mass.
Owing to the large uncertainties on the determinations, the derived values of the mass of NGC\,6705 are in reasonable agreement with the virial masses derived in Sect.~\ref{virmass}.

\begin{table*}
\begin{center}
	\caption{ \label{tab:mass} Mass contained in different regions of the cluster.}
	\small\addtolength{\tabcolsep}{-1pt}
	\begin{tabular}{l l l l l l}
	  \hline
	  \hline
	  Region & \multicolumn{1}{c}{$\alpha$}  & Observed & Extrapolated (power-law) & Extrapolated (Kroupa)\\
	  \multicolumn{1}{c}{[$\arcmin$]} &  & \multicolumn{1}{c}{[M$_{\sun}$]} & \multicolumn{1}{c}{[M$_{\sun}$]} & \multicolumn{1}{c}{[M$_{\sun}$]}\\ 
	  \hline
	  0 -- 3 & 1.18 $\pm$ 0.11 & 242 $\pm$ 64 & 375 $\pm$ 56     & 664  $\pm$ 115 \\
	  3 -- 6 & 3.01 $\pm$ 0.11 & 241 $\pm$ 46 & 4303 $\pm$ 1127  & 1430 $\pm$ 151 \\
	  6 -- 9 & 3.79 $\pm$ 0.10 & 181 $\pm$ 43 & 18\,888 $\pm$ 4702 & 1437 $\pm$ 242 \\
	  Total &                 & 664 $\pm$ 153 & 26\,566 $\pm$ 5885 & 3531 $\pm$ 508 \\ 
	  \hline
	  0 -- 9 & 2.70 $\pm$ 0.19 & 700 $\pm$ 310 & 6851 $\pm$ 1865  & 3683 $\pm$ 1063\\
	  \hline
	\end{tabular}
\tablefoot{$\alpha$ is the parameter in the mass function: $N(M)\,dM\propto M^{-\alpha}$. The stars were observed down to 1\,M$_{\sun}$. The extrapolation was done down to 0.08\,M$_{\sun}$.}
\end{center}
\end{table*}

\section{Chemical  analysis of the red clump stars} \label{sec:homogeneity}

\begin{figure}[ht]
\begin{center} \resizebox{\hsize}{!}{\includegraphics[scale=0.7]{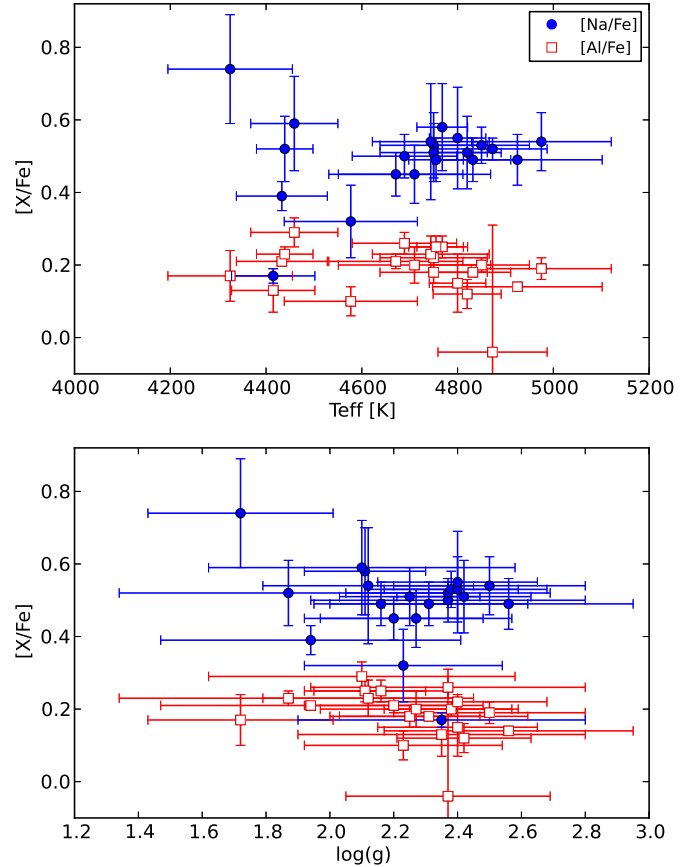}}
 \caption{\label{fig:teff}Solar-scaled abundances of [Na/Fe] (blue filled dots) and [Al/Fe] (red open squares) for the 21 UVES member stars, as a function of T$_{\textrm{eff}}$ and
log~$g$.}
 \end{center}
\end{figure}

The average iron abundance of the bona-fide members from UVES spectroscopy is [Fe/H]=0.10 with a dispersion of 0.06. 
This value is in agreement with \citet{GW00} who found iron abundances between 
0.07 and 0.20 from high-resolution spectroscopy of 10 K-giants of NGC\,6705 and with the values we derived from isochrone fitting.  In this section
we  study the relations between the abundances of Al and of other proton-capture elements Na, Mg, and 
Si.
 Al and Na abundances are plotted against T$_{\textrm{eff}}$ and
log~$g$ in  Fig.~\ref{fig:teff}. These abundances
do not display any significant trend. We observe
 star-to-star variations in the Al and Na content of the order of 0.2 dex, inside the expected
uncertainties. Similar diagnostics for Mg and Si are shown in \citet{Magrini14}.

\begin{figure*}[ht]
\begin{center} \includegraphics[scale=0.6]{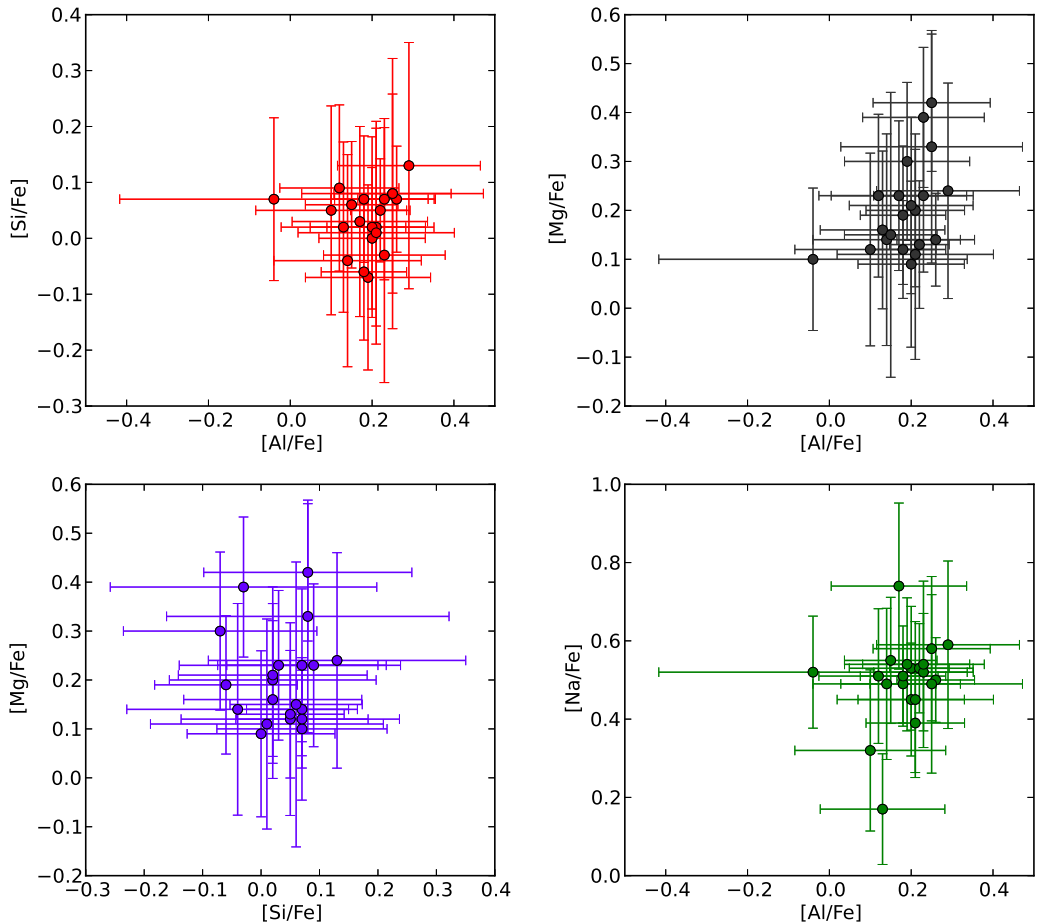} \caption{\label{fig:correlations}Abundances of Al, Si, Mg and Na for the 21 red clump stars. None of the "classical" correlations or anti-correlations between elemental abundances are observed in this cluster, which seems chemically homogeneous within our uncertainties.} \end{center}
\end{figure*}

Abundances of Al, Mg, Si and Na are plotted in Fig.~\ref{fig:correlations}. In case of multiple populations, correlations between the abundances of Al and Si, and between Al and Na, and anti-correlations between Si and Mg and between Al and Mg \citep{Gratton12} can be found. 
Models predict that Al should show correlations with elements that are
enhanced by the action of the Ne–Na (such as Na) and Mg–Al
cycles, and is anti-correlated with elements that are depleted in
H burning at high temperature (such as O and Mg).
We do not see any sign of such (anti-)correlations within our uncertainties.

One star shows a low Al abundance with a large uncertainty, with [Al/Fe]$=-0.04\pm0.38$, probably due to the low number of detected lines for this element. Two outliers in Na abundance could suggest an internal spread, but they are still compatible with a homogeneous cluster within the uncertainties. The star with the highest Na abundance ([Na/Fe]=+0.74\,dex) has the lowest Fe abundance in the sample ([Fe/H]=+0.03\,dex), while the star with the lowest Na content ([Na/Fe]=+0.17\,dex) has a Fe content [Fe/H]=+0.18\,dex. This suggests that the apparent discrepant [Na/Fe] for that star can be explained by considering the uncertainties on both Na and Fe abundances.

The mean [Na/Fe] for the cluster members is 0.48\,dex. This value is significantly higher than the Na abundance derived in field stars.
\citep{Soubiran05} show that at high metallicities thin disk stars exhibit an average of [Na/Fe]=0.11\,dex, even though they notice a rise at super-solar metallicities. 
For all elements presented here, the abundances were calculated in the LTE approximation. Na is subject to deviations from LTE \citep[e.g.,][]{Lind11}. The effect of non-LTE correction in stars of super-solar metallicity is expected to be small, but could lower the Na abundance of our sample by 0.15--0.20\,dex at most. On a side note, since the corrections depend on the evolutionary stage of the star, we expect them to affect all of the red clump stars in the same way, and not to create an additional spread in Na.

The approach presented in Sect.~\ref{sec:intrinsic} to estimate the intrinsic radial velocity dispersion of the sample can also be applied to the elemental abundances. Assuming an intrinsic Gaussian distribution of the abundance for each element, the most probable mean abundance ($\mu$) and intrinsic dispersion ($\sigma$) we computed are listed in Table~\ref{tab:abundancesM11}. For all five elements, the observed distribution is compatible with the cluster being homogeneous, and the observed spread can be entirely attributed to the individual uncertainties.

As far as we can see, NGC\,6705 is clearly a homogeneous object where the star-to-star scatter
is explained by the  uncertainties on the determination of the chemical abundances.
On the observational ground, 
\citet{Carretta10} suggested that clusters above a certain mass limit would develop multiple populations,
and derive the minimum (present-day) threshold mass of about $4 \times 10^{4}$\,M$_{\odot}$.
On the theoretical ground, by comparing the timescale of the mixing of heavy elements in protoclusters with the cluster 
formation timescale, \citet{Bland-Hawthorn10} suggest that  clusters up to $\sim 10^4$\,M$_{\odot}$ masses,
 and possibly a significant fraction of those as massive as $\sim 10^5$ M$_{\odot}$,
 should be chemically homogeneous.
As we have seen, the mass of NGC\,6705 is lower than $10^4$\,M$_{\odot}$ (see Sect.~\ref{massfunct}).
This mass is still well under the expected $4 \times 10^{4}$\,M$_{\odot}$ mass threshold. Figure~3 in \citet{Carretta10} \citep[updated in][]{Bragaglia13} shows that the least massive GC known to present multiple populations is Pal~5, with a mass of $6.3 \times 10^{4}$\,M$_{\odot}$. Terzan~7 and Pal~12, with masses of $3.9 \times 10^{4}$\,M$_{\odot}$ and $2.8 \times 10^{4}$\,M$_{\odot}$ (respectively) do not show multiple populations. 
Due to its mass, the chemical homogeneity of NGC\,6705 and the lack of the element correlations expected in 
clusters showing multiple populations, is consistent with the mass being the driving factor of cluster chemical evolution. However, the properties of M\,11 differ in several respects from most GCs: it is younger, has a supra-solar metallicity (it is much more metal-rich than Pal\,5, the least massive GCs known to contain multiple populations) and likely formed in a very different environment. While the lack of chemical inhomogeneities in M\,11 confirms the above scenario, it cannot completely rule out alternative possibilities. However, this result needs to be verifie on a larger sample.

\section{Summary}

We have used wide-field photometry from the ESO WFI and from the VPHAS+ Survey together with spectroscopic data derived in the framework of the GES to study the inner-disk cluster NGC\,6705.
We derive the extinction and the extinction law in front of the cluster. 
While the ratio $E(V-I)/ E(B-V)$ is consistent with the total-to selective extinction ratio R$_V=A_V/E(B-V)$=3.1
using the \citet{Fitzpatrick99} law, the relation $E(r-i)/ E(B-V)$ is slightly higher than the value of
 $E(r-i)=0.6 \times E(B-V)$ given by \citet{Yuan13}.
Using the red clump stars as tracers, we derive the extinction for the distance range 2 -- 4\,Kpc, where the cluster
is located.
The inner region of the cluster studied with the WFI $BVI$ photometry sits in a zone of relatively low and uniform extinction, with $E(B-V)=0.40$ out to $11\arcmin$ from the center. A higher extinction is found in the north-west quadrant,  with $E(B-V)$ up to 0.7.

We attribute membership probabilities to the 1028 main-sequence stars for which radial velocities were available from the GES.  We estimate the age of the cluster between 250 and 316\,Myr depending on the adopted set of
stellar isochrones. Its distance modulus turns out to be  11.45 $\pm$ 0.2 (d$_{\sun}=$1950 $\pm$ 200\,pc). By fitting a two-parameter King profile, we find a core radius of $1.23 \pm 0.28\arcmin$ (0.69 $\pm$ 0.23\,pc). The brightest stars of the cluster show a tighter spatial distribution, while the faintest stars are more widely distributed. Studying the mass function in different regions of the cluster shows that the inner parts contain a larger proportion of massive stars than the outer parts, confirming  that NGC\,6705 is undergoing mass segregation.
Using the velocity dispersion of the sample of members stars observed by the GES consortium using UVES, we derive the virial mass of
M$_{tot}=5083\pm1600$\,M$_{\odot}$, in agreement with the virial mass of $5621$\,M$_{\odot}$ found by \citet{McNamara77} using proper motions. 
The sample of stars observed by GIRAFFE over a larger radius presents  considerably higher velocity dispersion (4.52$\arcmin$ and 2.8\,km\,s$^{-1}$, respectively). This sample leads to a virial mass of 21\,900$\pm$2700\,M$_{\odot}$. However, the uncertainties on the GIRAFFE RVs are most likely underestimated, and this high virial mass must be considered an upper limit.
We estimated a mass from photometry, converting the $V$-band luminosity function to a mass function. When integrating the mass function of the whole 0-9$\arcmin$ region, we obtain values between 3683 $\pm$ 1063\,M$_{\sun}$ (using the Kroupa IMF under 1\,M$_{\sun}$) and 6851 $\pm$ 1865\,M$_{\sun}$ (using the extrapolated power-law).

The GES data enables us to study the chemical composition of the red clump stars. We find an iron abundance [Fe/H]=0.10 with a dispersion of 0.06. Despite being a massive open cluster, NGC\,6705 does not show any sign of an inhomogeneous stellar population of the kind observed in globular clusters. 
This cluster is well below  the mass limit of $4 \times 10^{4}$\,M$_{\odot}$ proposed in literature for clusters hosting multiple populations.  However, few massive clusters have been observed with high-resolution spectroscopy, and more studies are needed in order to observationally confirm this mass threshold. 

The reliability of the results that can be obtained for stellar clusters greatly depends on the determination of the membership and chemical abundances in the inner regions and in the outskirts, where haloes of unbound stars can be present. 
This especially holds for open clusters, which are more affected by the tidal field of the disks and by background contamination. The synergy between the GES and the ground-based surveys in general and the Gaia mission will provide invaluable information on the chemical abundances, distances and proper motions  of unbiased samples of stars, allowing for more accurate determinations of the cluster properties.

\section*{Acknowledgements}
Based on data products from observations made with ESO Telescopes at the La Silla Paranal Observatory under programme ID 188.B-3002.
This work was partially supported by the Gaia Research for European Astronomy Training (GREAT-ITN) Marie Curie network, funded through the European Union Seventh Framework Programme (FP7/2007-2013) under grant agreement no. 264895, partly supported by the European Union FP7 programme through ERC grant number 320360 and by the Leverhulme Trust through grant RPG-2012-541.
We acknowledge the support from INAF and Ministero dell’ Istruzione, dell'Universit\`a' e della Ricerca (MIUR) in the form of the grants "Premiale VLT 2012" and "The Chemical and Dynamical Evolution of the Milky Way and Local Group Galaxies" (prot. 2010LY5N2T).

The results presented here benefit from discussions held during the Gaia-ESO workshops and conferences supported by the ESF (European Science Foundation) through the GREAT Research Network Programme.
This work was supported in part by the National Science Foundation under Grant No. PHYS-1066293 and the hospitality of the Aspen Center for Physics during summer 2013.

The authors wish to thank M. Bellazzini for his suggestions and comments that helped improve this work.

L.S. aknowledges the support of Project IC120009 "Millennium Institute of Astrophysics (MAS)" of Iniciativa Científica Milenio del Ministerio de Economía, Fomento y Turismo de Chile.

T.B. was funded by grant No. 621-2009-3911 from The Swedish Research Council.
I.S.R. gratefully acknowledges the support provided by the Gemini-CONICYT project 32110029.
S.V. gratefully acknowledges the support provided by FONDECYT N. 1130721.


\begin{thebibliography}{99}

\bibitem[\protect\citeauthoryear{Allison et al.}{2009}]{Allison09} Allison, R. J., Goodwin, S. P., Parker, R. J., et al. 2009, MNRAS, 395, 1449

\bibitem[\protect\citeauthoryear{Bastian \& Silva-Villa}{2013}]{Bastian13} Bastian, N., Silva-Villa, E. 2013, MNRAS, 431, L122 

\bibitem[\protect\citeauthoryear{Bastian et al.}{2013}]{Bastian13b} Bastian, N., Lamers, H.~J.~G.~L.~M., de Mink, S.~E., et al. 2013, MNRAS, 436, 2398


\bibitem[\protect\citeauthoryear{Bellazzini et al.}{2012}]{Bellazzini12}
{Bellazzini}, M., {Bragaglia}, A., {Carretta}, E., {Gratton}, R.~G., {Lucatello}, S., {Catanzaro}, G., {Leone}, F. 2012, A\&A, 493, 959



\bibitem[\protect\citeauthoryear{Bianchini et al.}{2013}]{Bianchini13}
{Bianchini}, P., {Varri}, A.~L., {Bertin}, G., {Zocchi}, A. 2013, ApJ, 772, 67


\bibitem[\protect\citeauthoryear{Bland-Hawthorn et al.}{2010}]{Bland-Hawthorn10}
{Bland-Hawthorn}, J., {Karlsson}, T., {Sharma}, S., {Krumholz}, M. and {Silk}, J. 2010, ApJ, 721, 582

\bibitem[\protect\citeauthoryear{Bragaglia et al.}{2012}]{Bragaglia12} Bragaglia, A., Gratton, R. G., Carretta, E., et al. 2012, A\&A, 548, A122

\bibitem[\protect\citeauthoryear{Bragaglia et 
al.}{2013}]{Bragaglia13} Bragaglia, A., Sneden, C., Gratton. R.G., Carretta, E., Lucatello, S. 2013, \apj, submitted



\bibitem[\protect\citeauthoryear{Bressan et al.}{2012}]{Bressan12}
Bressan, A., Marigo, P., Girardi, L., et al. 2012, MNRAS, 427, 127



\bibitem[\protect\citeauthoryear{Cabrera-Ca\~no \& Alfaro}{1985}]{Cabrera85} Cabrera-Cano J., Alfaro E.~J., 1985, A\&A, 150, 298 

\bibitem[\protect\citeauthoryear{Carretta et al.}{2009a}]{Carretta09a} Carretta, E., Bragaglia, A., Gratton, R.~G., et al.\ 2009, \aap, 505, 117 

\bibitem[\protect\citeauthoryear{Carretta et al.}{2009b}]{Carretta09b} Carretta, E., Bragaglia, A., Gratton, R.~G., \& Lucatello, S.\ 2009, \aap, 505, 139 

\bibitem[\protect\citeauthoryear{Carretta et al.}{2010}]{Carretta10}
Carretta, E., Bragaglia, A., Gratton, R. G., et al. 2010, A\&A, 516, A55

\bibitem[\protect\citeauthoryear{Carretta et 
al.}{2013}]{Carretta13} Carretta, E., Bragaglia, A., Gratton, R.~G., et al. 2013, A\&A, 561, A87

\bibitem[\protect\citeauthoryear{Cote et al.}{1995}]{Cote95}
{Cote}, P., {Welch}, D.~L., {Fischer}, P., {Gebhardt}, K. 1995, ApJ, 454, 788

\bibitem[\protect\citeauthoryear{Decressin et al.}{2007}]{Decressin07} Decressin, T., Meynet, G., Charbonnel, C., Prantzos, N., \& Ekstr{\"o}m, S.\ 2007, \aap, 464, 1029 

\bibitem[\protect\citeauthoryear{D'Ercole et al.}{2008}]{DErcole08} D'Ercole, A., Vesperini, E., D'Antona, F., McMillan, S.~L.~W., \& Recchi, S.\ 2008, \mnras, 391, 825 

\bibitem[\protect\citeauthoryear{Donati et al.}{2012}]{Donati12}
Donati, P., Bragaglia, A., Cignoni, M., Cocozza, G., Tosi, M. 2012, MNRAS, 424, 1132

\bibitem[\protect\citeauthoryear{D'Orazi et al.}{2010}]{DOrazi10}
D’Orazi, V., Lucatello, S., Gratton, R., et al. 2010, ApJ, 713, L1

\bibitem[\protect\citeauthoryear{Drew et al.}{2014}]{Drew14}
Drew, J.~E., Gonzalez-Solares, E., Greimel, R. et al. 2014, MNRAS, 440, 2036



\bibitem[\protect\citeauthoryear{Dotter et al.}{2008}]{Dotter08}
Dotter, A., Chaboyer, B., {Jevremovi{\'c}}, D., et al. 2008, ApJS, 178, 89



\bibitem[\protect\citeauthoryear{Fitzpatrick}{1999}]{Fitzpatrick99}
Fitzpatrick E. L., 1999, PASP, 111, 6

\bibitem[\protect\citeauthoryear{Geisler et al.}{2012}]{Geisler12}
Geisler, D., Villanova, S., Carraro, G., et al. 2012, ApJ, 756, L40

\bibitem[\protect\citeauthoryear{Gilmore et al.}{2012}]{Gilmore12}
Gilmore, G., Randich, S., Asplund, M., et al. 2012, The Messenger, 147, 25

\bibitem[\protect\citeauthoryear{Gonzalez \& Wallerstein}{2000}]{GW00}
Gonzalez, G., Wallerstein, G. 2000, PASP, 112, 1081

\bibitem[\protect\citeauthoryear{Gratton et 
al.}{2001}]{Gratton01} Gratton, R.~G., Bonifacio, P., Bragaglia, A., et al.\ 2001, \aap, 369, 87 

\bibitem[\protect\citeauthoryear{Gratton et al.}{2004}]{Gratton04} Gratton, R., Sneden, C., \& Carretta, E.\ 2004, \araa, 42, 385 

\bibitem[\protect\citeauthoryear{Gratton et al.}{2012}]{Gratton12} Gratton, R.~G., Carretta, E., \& Bragaglia, A.\ 2012, \aapr, 20, 50 

\bibitem[\protect\citeauthoryear{Grevesse et al.}{2007}]{Grevesse07}
Grevesse, N., Asplund, M., Sauval, A. J. 2007, SSRv,130, 105

\bibitem[\protect\citeauthoryear{Gustafsson et al.}{2008}]{MARCS} Gustafsson, B., Edvardsson, B., Eriksson, K., et al., 2008, A\&A, 486, 951 

\bibitem[\protect\citeauthoryear{Heiter et al.}{in prep.}]{Heiter14} Heiter, U., and the GES Line List group, in preparation.

\bibitem[\protect\citeauthoryear{Jofre et al.}{2013}]{Jofre13} Jofre P., Heiter U., Soubiran C. et al. 2013, arXiv:1309.1099

\bibitem[\protect\citeauthoryear{Johnson \& Pilachowski}{2010}]{Johnson10} Johnson, C.~I., \& Pilachowski, C.~A.\ 2010, \apj, 722, 1373 

\bibitem[\protect\citeauthoryear{King}{1962}]{King62}
King, I. 1962, AJ, 67, 471

\bibitem[\protect\citeauthoryear{Koo et al.}{2007}]{Koo07}
Koo, J.-R., Kim, S.-L., Rey, S.-C., et al. 2007, PASP, 119, 1233

\bibitem[\protect\citeauthoryear{Kraft}{1979}]{Kraft79} Kraft, R.~P.\ 1979, \araa, 17, 309 

\bibitem[\protect\citeauthoryear{Kraft}{1994}]{Kraft94} Kraft, R.~P.\ 1994, \pasp, 106, 
553 

\bibitem[\protect\citeauthoryear{Kroupa et al.}{2001}]{Kroupa01}
Kroupa, P. 2001, MNRAS, 322, 231

\bibitem[\protect\citeauthoryear{Lee et al.}{1999}]{Lee99} Lee, Y.-W., Joo, J.-M., Sohn, Y.-J., Rey, S.-C., Lee, H.-C., Walker, A.R. 1999, Nature 402, 55

\bibitem[\protect\citeauthoryear{Lind et 
al.}{2009}]{Lind09} Lind, K., Primas, F., Charbonnel, C., Grundahl, F., \& Asplund, M.\ 2009, \aap, 503, 545 

\bibitem[\protect\citeauthoryear{Lind et 
al.}{2011}]{Lind11} Lind K., Asplund M., Barklem P.~S., Belyaev A.~K., 2011, A\&A, 528, A103 

\bibitem[\protect\citeauthoryear{Lovis \& Mayor}{2007}]{Lovis07}
Lovis, C., Mayor, M. 2007, A\&A, 472, 657

\bibitem[\protect\citeauthoryear{Magrini et al.}{2009}]{Magrini09}
Magrini, L., Sestito, P., Randich, S., \& Galli, D. 2009, A\&A, 494, 95

\bibitem[\protect\citeauthoryear{Magrini et al.}{2010}]{Magrini10}
Magrini, L., Randich, S., Zoccali, M., et al. 2010, A\&A, 523, A11

\bibitem[\protect\citeauthoryear{Magrini et al.}{2014}]{Magrini14}
Magrini, L., Randich, S., Romano, D., et al. 2014, A\&A, 563, 44

\bibitem[\protect\citeauthoryear{Marino et 
al.}{2008}]{Marino08} Marino, A.~F., Villanova, S., Piotto, G., et al. 2008, \aap, 490, 625 

\bibitem[\protect\citeauthoryear{Marino et 
al.}{2011}]{Marino11} Marino, A.~F., Sneden, C., Kraft, R.~P., et al. 2011, \aap, 532, A8 

\bibitem[\protect\citeauthoryear{Martell et al.}{2011}]{Martell11} Martell, S.~L., Smolinski, J.~P., Beers, T.~C., Grebel, E.~K. 2011, A\&A, 534, A136

\bibitem[\protect\citeauthoryear{Mathieu}{1984}]{Mathieu84}
Mathieu, R. D. 1984, ApJ, 284, 643

\bibitem[\protect\citeauthoryear{McNamara \& Sanders}{1977}]{McNamara77}
McNamara, B. J., Sanders, W. L. 1977, A\&A, 54, 569

\bibitem[\protect\citeauthoryear{Messina et al.}{2010}]{Messina10}
Messina, S., Parihar, P., Koo, J.-R., et al. 2010, A\&A, 513, A29

\bibitem[\protect\citeauthoryear{Milone et al.}{2012}]{Milone12} Milone, A.~P., Piotto, G., Bedin, L.~R., et al.\ 2012, \apj, 744, 58 

\bibitem[\protect\citeauthoryear{Miocchi}{2006}]{Miocchi06} Miocchi, P. 2006, MNRAS, 366, 227

\bibitem[\protect\citeauthoryear{Piotto}{2009}]{Piotto09} Piotto, G.\ 2009, IAU Symposium, 258, 233 

\bibitem[\protect\citeauthoryear{Randich \& Gilmore}{2013}]{Randich13} Randich, S., Gilmore, G. 2013, The Messenger, 154, 47

\bibitem[\protect\citeauthoryear{Ram{\'{\i}}rez 
\& Cohen}{2002}]{Ramirez02} Ram{\'{\i}}rez, S.~V., \& Cohen, J.~G.\ 2002, \aj, 123, 3277 

\bibitem[\protect\citeauthoryear{Robin et al.}{2003}]{Robin03}
Robin, A. C., {Reyl{\'e}}, C., {Derri{\`e}re}, S., Picaud, S. 2003, A\&A, 409, 523

\bibitem[\protect\citeauthoryear{Sacco et al.}{2014}]{Sacco14} Sacco, G.~G., Morbidelli, L., Franciosini, E. et al. 2014, A\&A, 565, 113 

\bibitem[\protect\citeauthoryear{Santos et al.}{1990}]{Santos90}
Santos, Jr., J. F. C., Bica, E., Dottori, H. 1990, PASP, 102, 454

\bibitem[\protect\citeauthoryear{Santos et al.}{2005}]{Santos05}
Santos, Jr., J. F. C., Bonatto, C., Bica, E. 2005, A\&A, 442, 201

\bibitem[\protect\citeauthoryear{Schlegel}{1998}]{Schlegel98} Schlegel D.~J., Finkbeiner D.~P., Davis M., 1998, ApJ, 500, 525 

\bibitem[\protect\citeauthoryear{Soubiran \& Girard}{2005}]{Soubiran05} Soubiran, C., Girard, P. 2005, A\&A, 438, 139


\bibitem[\protect\citeauthoryear{Spina et al.}{submitted}]{Spina14} Spina L. et al., submitted. 

\bibitem[\protect\citeauthoryear{Spitzer}{1969}]{Spitzer69}
Spitzer L. Jr., 1969, ApJ, 158, L139

\bibitem[\protect\citeauthoryear{Stetson}{1987}]{Stetson87} Stetson P.~B., 1987, PASP, 99, 191

\bibitem[\protect\citeauthoryear{Sung et al.}{1999}]{Sung99}
Sung, H., Bessell, M. S., Lee, H.-W., Kang, Y. H., Lee, S.-W. 1999, MNRAS, 310, 982

\bibitem[\protect\citeauthoryear{Vallenari et al.}{2006}]{PadovaGalMod}
Vallenari, A., Pasetto, S., Bertelli, G., et al. 2006, A\&A, 451, 125

\bibitem[\protect\citeauthoryear{Vandame}{2002}]{Vandame02}
Vandame, B. 2002, in Society of Photo-Optical Instrumentation Engineers
(SPIE) Conference Series, Vol. 4847, Society of Photo-Optical Instrumentation Engineers (SPIE) Conference Series, ed. J.-L. Starck \& F. D. Murtagh, 123–134

\bibitem[\protect\citeauthoryear{Ventura et al.}{2001}]{Ventura01} Ventura, P., D'Antona, 
F., Mazzitelli, I., \& Gratton, R.\ 2001, \apjl, 550, L65

\bibitem[\protect\citeauthoryear{Villanova et al.}{2013}]{Villanova13} Villanova, S., 
Geisler, D., Carraro, G., Moni Bidin, C., Munoz, C.\ 2013, A\&A, 778, 186 

\bibitem[\protect\citeauthoryear{Walker et al.}{2006}]{Walker06} Walker, M.~G., Mateo, M., Olszewski, E.~W., et al. 2006, AJ, 131, 2114 

\bibitem[\protect\citeauthoryear{Weidner et al.}{2009}]{Weidner09} Weidner, C., Kroupa, P., Maschberger, T., 2009, MNRAS, 393, 663 

\bibitem[\protect\citeauthoryear{Yuan et al.}{2013}]{Yuan13} Yuan H.~B., Liu X.~W., Xiang M.~S., 2013, MNRAS, 430, 2188 

\bibitem[\protect\citeauthoryear{Zacharias et al.}{2012}]{UCAC4} Zacharias, N., Finch, C. T., Girard, T. M., et al. 2012, VizieR Online Data Catalog: I/322 

\end{thebibliography}
\end{document}